\documentclass{emulateapj}
\usepackage{natbib} 
\usepackage{placeins}
\usepackage{rotate}
\usepackage{longtable}
\usepackage{threeparttable}
\usepackage{epsfig,graphicx}
\usepackage{float}
\usepackage[pass]{geometry}
\def\lap{\lower.5ex\hbox{$\; \buildrel < \over \sim \;$}}
\def\gap{\lower.5ex\hbox{$\; \buildrel > \over \sim \;$}}

\def\ergcm2s{${\rm erg\ cm^{-2}\ s^{-1}}$}

\def\ergscm2s{${\rm erg\ cm^{-2}\  s^{-1}}$}

\def\cm-2{${\rm cm^{-2}}$}

\begin{document}

\title{Comparing Chandra and Hubble in the Northern Disk of M31}

\author{Benjamin F. Williams\altaffilmark{1},
Margaret Lazzarini\altaffilmark{1},
Paul Plucinsky\altaffilmark{2},
Manami Sasaki\altaffilmark{3},
Vallia Antoniou\altaffilmark{2},
Neven Vulic\altaffilmark{4,5},
Michael Eracleous\altaffilmark{6},
Knox S. Long\altaffilmark{7,8},
Breanna Binder\altaffilmark{9},
Julianne Dalcanton\altaffilmark{1},
Alexia R. Lewis\altaffilmark{10},
Daniel R. Weisz\altaffilmark{11}
}

\altaffiltext{1}{Department of Astronomy, Box 351580, University of Washington, Seattle, WA 98195; ben@astro.washington.edu, mlazz@uw.edu,jd@astro.washington.edu}
\altaffiltext{2}{Harvard-Smithsonian Center for Astrophysics, MS-3, 60 Garden Street, Cambridge, MA, 02138; pplucinsky@cfa.harvard.edu}
\altaffiltext{3}{Friedrich-Alexander-Universit{\"a}t Erlangen-N{\"u}rnberg, Sternwartstrasse 7, 96049 Bamberg, Germany; manami.sasaki@fau.de}
\altaffiltext{4}{Laboratory for X-ray Astrophysics, NASA Goddard Space Flight Center, Code 662, Greenbelt, MD 20771}
\altaffiltext{5}{Department of Astronomy and Center for Space Science and Technology (CRESST), University of Maryland, College Park, MD 20742-2421, USA; neven.vulic@nasa.gov}
\altaffiltext{6}{Department of Astronomy and Astrophysics and Institute for Gravitation and the Cosmos, 525 Davey Lab, University Park, PA 16802.}
\altaffiltext{7}{Space Telescope Science Institute, Baltimore, MD 21218; long@stsci.edu}
\altaffiltext{8}{Eureka Scientific, Inc., 2452 Delmer Street, Suite 100, Oakland, CA 94602-3017, USA}
\altaffiltext{9}{Department of Physics and Astronomy, California State Polytechnic University, 3801 West Temple Avenue, Pomona, CA 91768; babinder@cpp.edu}
\altaffiltext{10}{Department of Astronomy, The Ohio State University, 140 West 18th Avenue, Columbus, OH 43210; lewis.1590@osu.edu}
\altaffiltext{11}{Department of Astronomy, University of California Berkeley, Berkeley, CA 94720; dan.weisz@berkeley.edu}

\begin{abstract}
The X-ray source populations within galaxies are typically difficult
to identify and classify from X-ray data alone.  We are able to break
through this barrier by combining deep new Chandra ACIS-I observations
with extensive Hubble Space Telescope imaging from the Panchromatic
Hubble Andromeda Treasury of the M31 disk. We detect 373 X-ray
  sources down to 0.35-8.0 keV flux of 10$^{-15}$ erg cm$^{-2}$
  s$^{-1}$ over 0.4 square degrees, 170 of which are reported for the
first time.  We identify optical counterpart candidates for 188 of the
373 sources, after using the HST data to correct the absolute
astrometry of our Chandra imaging to 0.1$''$.  While 58 of these 188
are associated with point sources potentially in M31, over half (107)
of the counterpart candidates are extended background galaxies, 5 are
star clusters, 12 are foreground stars, and 6 are supernova remnants.
Sources with no clear counterpart candidate are most likely to be
  undetected background galaxies and low-mass X-ray binaries in M31.
The hardest sources in the $1-8$~keV band tend to be matched to
background galaxies. The 58 point sources that are not consistent with
foreground stars are bright enough that they could be high mass stars
in M31; however, all but 8 have optical colors inconsistent with
single stars, suggesting that many could be background galaxies or
binary counterparts. For point-like counterparts, we examine the star
formation history of the surrounding stellar populations to look for a
young component that could be associated with a high mass X-ray binary
(HMXB).  About one third of the point sources are not physically
associated with a young population, and are therefore more likely to
be background galaxies.  For the 40 point-like counterpart candidates
associated with young populations, we find that their age distribution
has two peaks at 15-20 Myr and 40-50 Myr.  If we only consider the 8
counterpart candidates with typical high-mass main sequence optical
star colors, their age distribution peaks mimic those of the
  sample of 40.  Finally, we find that intrinsic faintness, and not
extinction, is the main limitation for finding further counterpart
candidates.

\end{abstract}

\section{Introduction}

X-ray sources probe the most exotic forms of matter in the universe.
Those outside of active galactic nuclei, such as X-ray
binaries (XRBs) and supernova remnants (SNRs), can only be detected in
nearby galaxies.  Chandra and XMM-Newton can resolve hundreds of
individual stellar-mass X-ray sources in Local Group galaxies, but
outside of the Magellanic Clouds, the identification of counterparts
for these stellar mass X-ray sources has been hampered by low spatial
resolution X-ray data, difficulty separating background galaxies from
stars in optical imaging, and stellar crowding.

Over the past decade, our ability to identify high-quality counterpart
candidates for X-ray sources outside of the Galaxy and Magellanic
Clouds has improved greatly due to the combination of high spatial
resolution X-ray imaging with Chandra and resolved stellar photometry
with the Hubble Space Telescope (HST).  Populations of OB star and
background galaxy counterpart candidates have been classified out to
distances of 3~Mpc, in particular, M31 \citep[770
  kpc;][]{williams2005a,
  williams2005b,williams2005c,hatzidimitriou2006,williams2014hmxbs},
NGC~300 \citep[2 Mpc;][]{binder2012}, NGC~2403 \citep[3
  Mpc;][]{binder2015}, and NGC~404 \citep[3 Mpc;][]{binder2013}.

As the nearest massive spiral, M31 has been observed extensively in
X-rays.  Building on early surveys with the Einstein observatory
\citep{vanspeybroeck1979} and ROSAT \citep{supper1997,supper2001},
XMM-Newton has mapped the entire interior of the $D_{25}$ isophotal
contour \citep[][hereafter S11]{pietsch2005,stiele2011}, and Chandra
has observed the inner disk with the HRC \citep{williams2004}, mapped
portions of the disk with ACIS \citep{distefano2004,vulic2016}, and
monitored the bulge and nuclear regions for over a decade
\citep{kong2002,kaaret2002,garcia2010,li2011}. These surveys have
detected dozens of transient X-ray sources, and thousands of
persistent sources.

Most of the known X-ray sources are unidentified, but many are
consistent with emission originating from background active galactic
nuclei (AGNs).  Others are clearly matched to bright Milky Way
foreground stars.  However, the most interesting sources are those
that may truly be in M31.  \citet{pietsch2005} and S11 provided
hundreds of source classifications based on variability and hardness
ratios, and they identify dozens of SNRs and XRB candidates based on
cross-matching with bright stars and star clusters from ground-based
imaging and catalogs.  With all of this work, only two strong high
mass X-ray binary (HMXB) candidates were seen (sources 1579 and 1716
in S11), potentially because of the difficulties of identifying
stellar counterparts in the crowded M31 with the spatial resolution
available in these data.  By comparing the XMM-Newton catalogs and
ground-based photometry, \citet{williams2014hmxbs} obtained spectra of
dozens of optical counterpart candidates in the M31 field, finding few,
if any HMXBs.  Most of their spectra showed the counterparts to be
background AGN, even though they were targeted to be blue point sources
in ground based imaging.

While all of this work has significantly advanced our knowledge of
M31's X-ray source populations, at this point it remains unclear what
fraction of the known X-ray sources actually belong to M31 and
  which of the sources are background galaxies being viewed through
  M31.  It is also unclear what the nature of most M31 sources
  is.

The very recent HST survey of the northern half of M31, the
    Panchromatic Hubble Andromeda Treasury
    \citep[PHAT;][]{dalcanton2012,williams2014}, offers an opportunity
    to remedy this situation.  PHAT is the largest HST mosaic ever
    assembled, covering a large fraction of the northern M31 disk in 6
    HST filters from the near ultraviolet to the near infrared,
    supplying photometry for over 100 million stars.  The high-resolution imaging provides the opportunity to find high-quality counterpart candidates for background galaxies and HMXBs.  The resolved photometry allows us to determine the physical characteristics of the stellar populations surrounding the X-ray sources.  

In addition to allowing us to optically identify background
  galaxies, we expect the PHAT \citep{dalcanton2012,williams2014}
footprint to contain $\gap$20 HMXBs. Measurements of the star
formation rate in M31 \citep{williams2003,lewis2015} are $\sim$0.3
M$_{\odot}$ yr$^{-1}$ in the PHAT footprint.  The scaling relation
between SFR and HMXBs \citep{grimm2003} implies $\sim$20 HMXBs with
L$_{\rm X}{>}10^{36}$ erg s$^{-1}$ should be associated with that
amount of star formation, and the relation scaling relation in the
Magellanic Clouds \citep{antoniou2010} suggests $\sim$100 Be-XRBs, but
only a fraction of these ($\sim$20) are expected to have high X-ray
luminosities, usually associated with accretion disk systems (i.e., fed by Roche-lobe overflow).
Combining the catalogs of \citet{binder2015} with the optical catalogs
from the ANGST program \citep{dalcanton2009} suggests a scaling
relation between the number of OB stars with M$_V{<}{-}1$ and the
number of bright HMXBs (L$_{\rm X}{>}10^{36}$ erg s$^{-1}$).  There
are $\sim$8$\times$10$^4$ such stars ($m_{f475w}{<}23.55$,
$m_{f475w}-m_{f814w}{<}0.5$) in the PHAT footprint, which implies
$\sim$30 bright HMXBs. Finally, the summed spectra of sources with
L$_{\rm X}{\sim}$(5--10)$\times$10$^{35}$ ergs s$^{-1}$ has a photon
index consistent with neutron star HMXBs \citep[see Figure 6 of]
[]{ShawGreening2009}, consistent with this estimate.

So, where are the bright HMXBs in M31?  Perhaps they are being missed
due to stellar crowding in the M31 disk.  To better localize the X-ray
sources, we have undertaken a Chandra survey covering much of the PHAT
survey area. We designed our survey to provide the largest number of
precise positions for the least amount of {\it Chandra} time. Our
final observations achieved a 0.35$-$8 keV depth of
3$\times$10$^{-15}$~erg~cm$^{-2}$~s$^{-1}$ (assuming a power-law
spectrum with an index of 1.7 and $N_H$=7$\times$10$^{20}$~cm$^{-2}$
(as in, e.g., S11), which corresponds to
$\sim$3$\times$10$^{35}$~erg~s$^{-1}$ at the distance of M31
\citep[770 kpc,][]{mcconnachie2005}.  This depth should allow us to
detect dozens of HMXBs and provide a reliable test of the predicted
numbers.

Low-mass X-ray binaries (LMXBs) are not as simple to identify, even with {\it HST imaging, as their optical counterparts are too faint to be distinguished.  However, based on the
    stellar mass maps from PHAT \citep{williams2017}, there are
    $\sim$2$\times$10$^{10}$~M$_{\odot}$ in the region covered by our
    X-ray data, suggesting a LMXB population with
    L$_X{>}$3$\times$10$^{35}$ erg s$^{-1}$ of $\sim$100 according to
    the LMXB X-ray luminosity function (XLF) from \citet{lehmer2014}.  By identifying a large fraction of the other sources, we can test this prediction for consistency.}

By combining Chandra positions with HST imaging, we simultaneously
limit the number of potential counterpart candidates, identify the
most likely counterpart to the X-ray source based on HST photometry,
and easily resolve many background galaxies, which we expect to
dominate the X-ray catalog.  For the best X-ray binary candidates, the
same data set can be used to constrain the progenitor age, physical
characteristics of the secondaries, and provide targets for follow-up
optical spectroscopy to measure orbital periods.  Ideally,
time-resolved spectroscopy of the resulting catalog will ultimately
provide clean age and orbital period distributions for a sample of M31
X-ray binaries, which can also be tied to the properties of their
local stellar populations.  Such a sample will provide quantitative
tests for predictions of HMXB production from binary evolution models.

In this paper, we present our initial catalog, counterpart candidates,
and measure the age distribution of HMXB candidates.  In Section 2, we
describe the observations of our Chandra survey of the PHAT region, as
well as our data reduction technique for measuring the X-ray sources
and aligning the Chandra data to PHAT directly.  In Section 3, we
present our Chandra catalog, cross-matched with the XMM-Newton catalog
of S11.  We include the most likely optical counterparts in cases
where a likely counterpart is present in the HST data.  In Section 4,
we describe some of the most interesting counterpart candidates,
including the best HMXB candidates, and in Section 5 we summarize our
work.

\section{Observations and Reductions}

\begin{turnpage}
\begin{figure*}
\begin{center}
\includegraphics[height=3.5in]{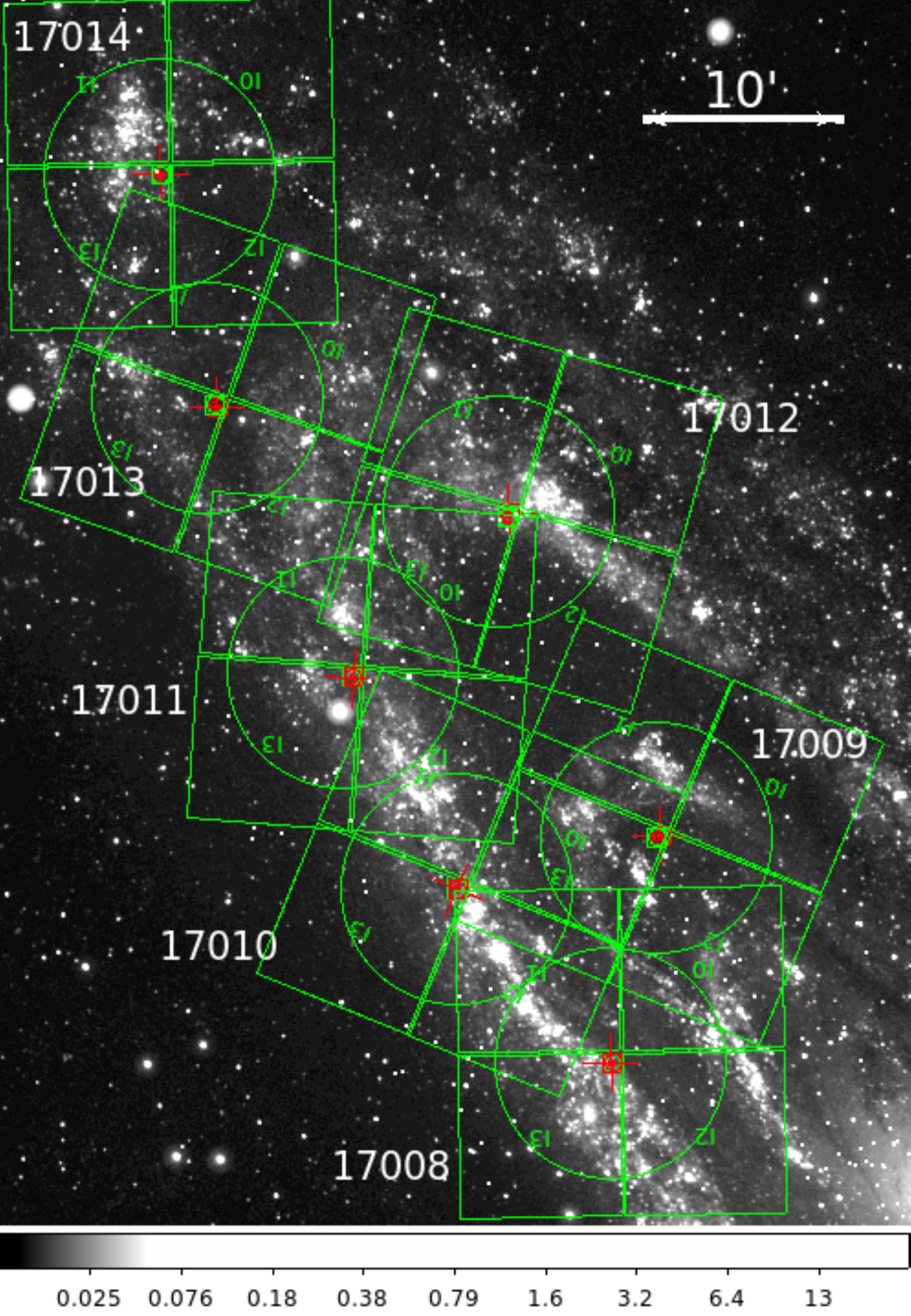}
\includegraphics[height=3.5in]{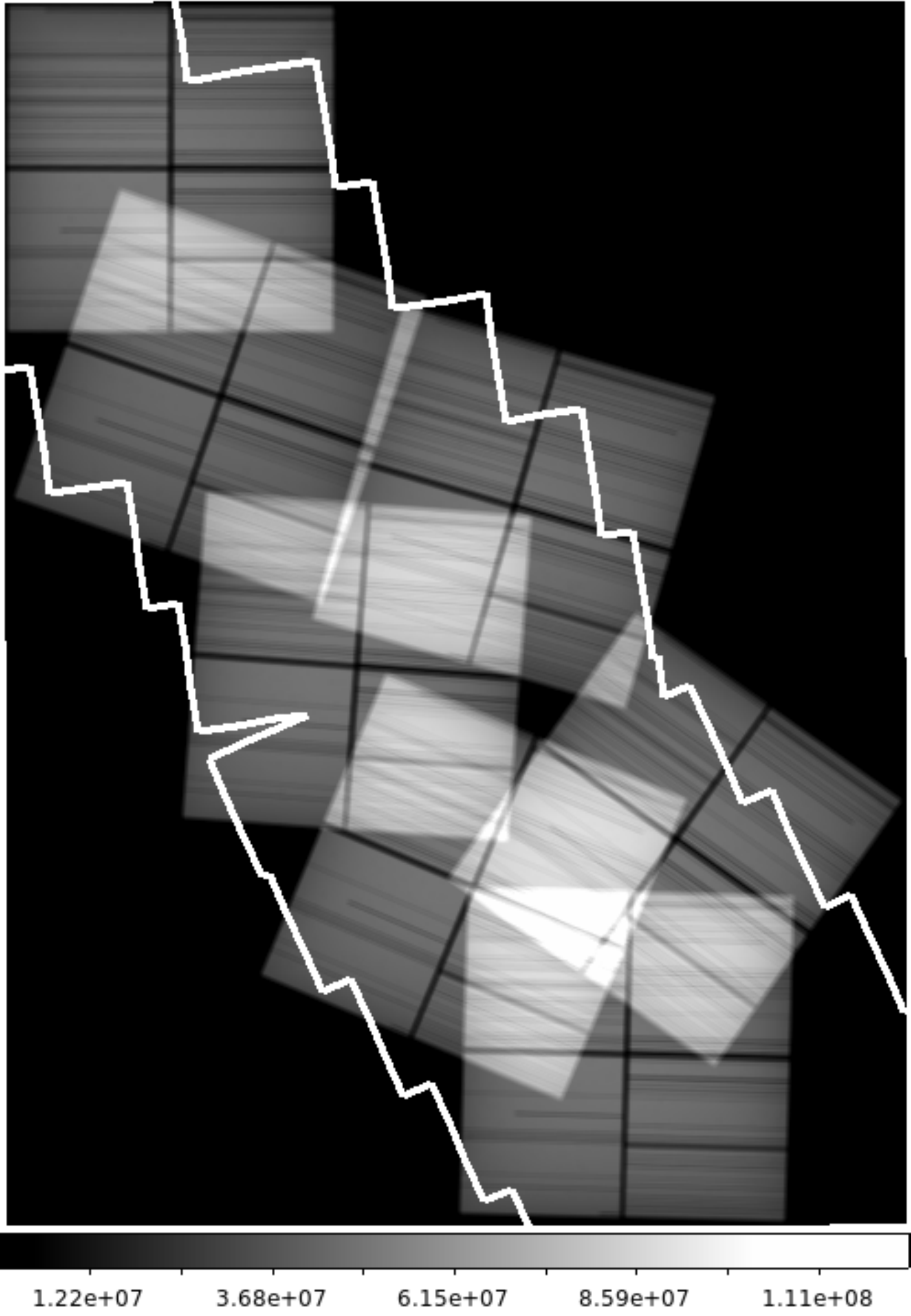}
\includegraphics[height=3.5in]{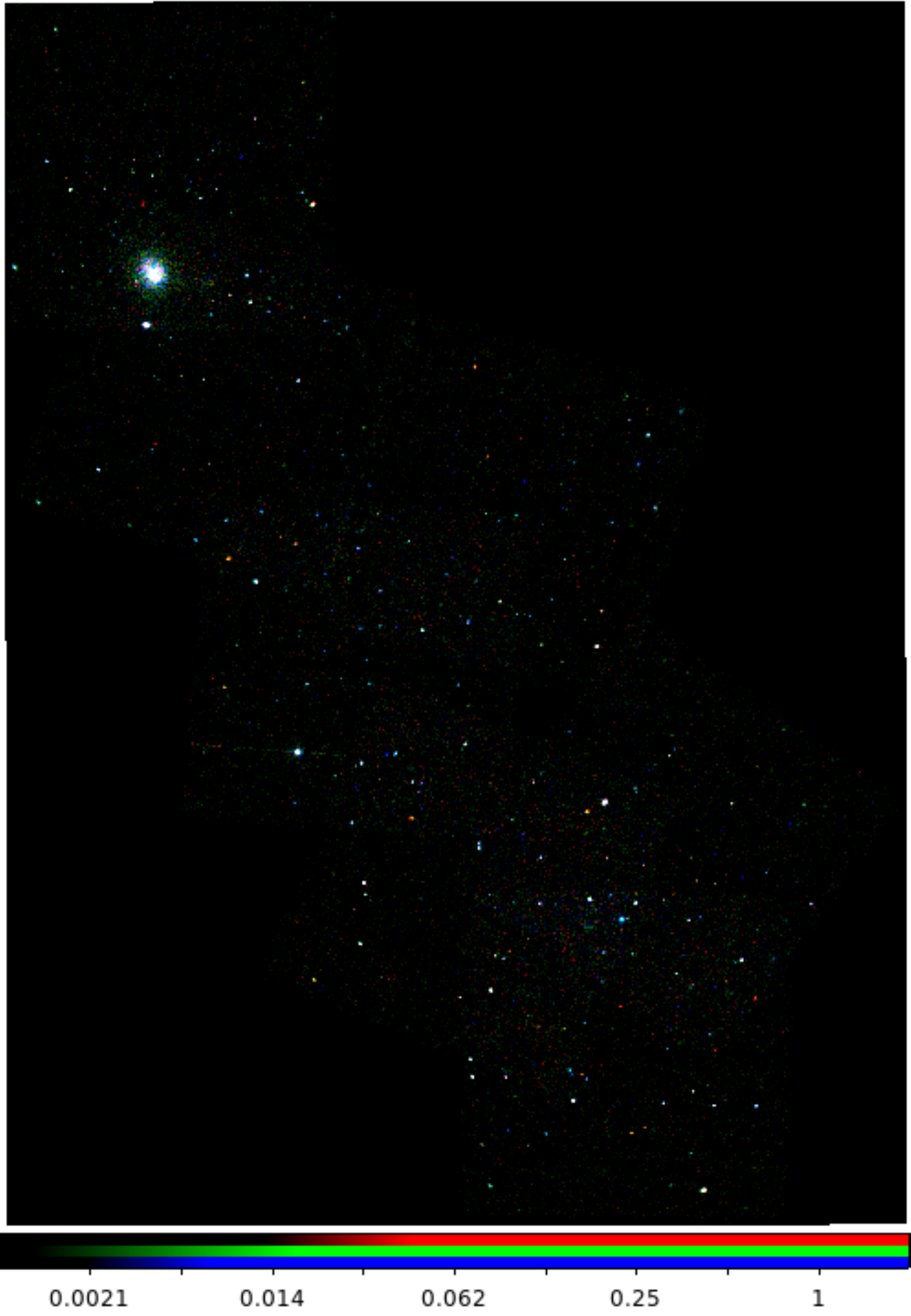}
\end{center}
\caption{{\it Left:} GALEX NUV image of M31 with our planned Chandra
  pointings overlaid.  This image is oriented with North-up,
  East-left, and the inset scale bar is 10$'$. {\it Center:} Exposure
  map of the Chandra survey as observed. The total coverage is 0.41 square
  degrees.  The white polygon marks the PHAT footprint, and the
  overlap is 0.36 square degrees. {\it Right:} Color mosaic of our
  Chandra data, where red is 0.3-1.0 keV, green is 1.0-2.0 keV, and
  blue is 2.0-8.0 keV. \label{footprints}}
\end{figure*}

\end{turnpage}

In October of 2015, we observed the PHAT footprint with Chandra with 7
pointings.  The footprints are overlaid on a GALEX NUV image of M31,
along with the corresponding HST coverage, in Figure~\ref{footprints}.
Using the S11 catalog, we found that by obtaining 25 counts for each
5$\times$10$^{-15}$~erg~cm$^{-2}$~s$^{-1}$ source, we would achieve
excellent efficiency in measuring positions, and we would detect
fainter sources (down to a faint limit of
  $\sim$10$^{-15}$~erg~cm$^{-2}$~s$^{-1}$) near the field
centers. Therefore, at each pointing we observed for about 50 ks in VF
mode.  These 7 observations are summarized in Table~\ref{obstab}.

\subsection{X-ray Photometry}

We processed these observations independently using CIAO 4.7 with
CALDB version 4.6.7 \citep{ciao}.  We first generated exposure maps
and images of the counts covering the full detector with 0.492$''$
pixel resolution using the task {\it fluximage}.  We made
corresponding maps of the point spread function using {\it mkpsfmap}.
These were put through the task {\it wavdetect} using the default
parameters, and searching on scales of 1, 2, 3, 8, and 16.  The output
source regions were then overplotted on the images and inspected
  by eye to assure that no obvious sources were missed and that all
sources appeared to correspond to true overdensities of counts.  At
the same time, we ensured that sources appearing in multiple
observations were noted so that they would appear in our catalog only
once, but their measurements in each observation were kept separate to
assess variability and position uncertainty. This process resulted in
a total of 373 unique detected sources.

\begin{figure*}
\begin{center}
\includegraphics[width=6.0in]{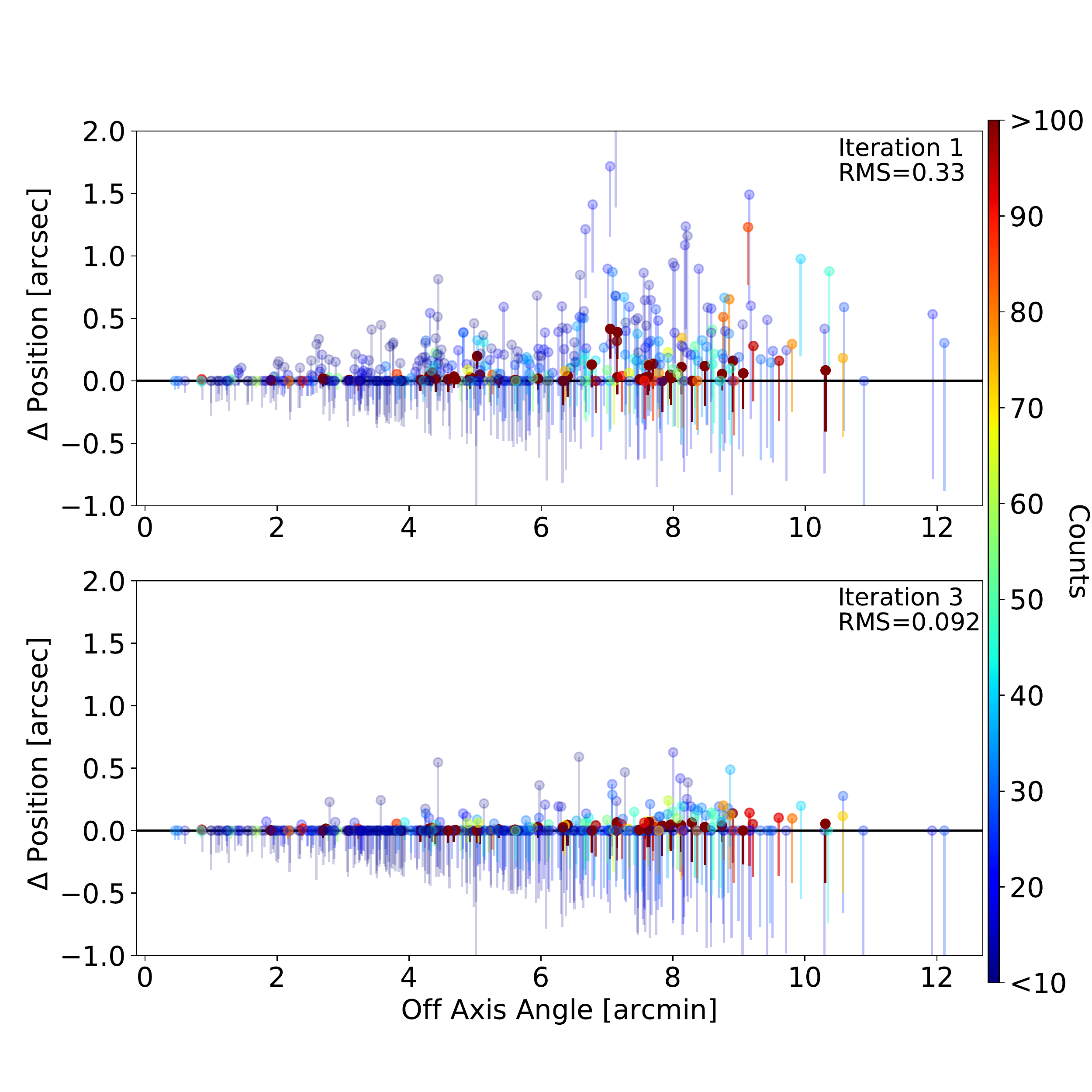}
\end{center}
\caption{Results of our AE {\tt fit\_positions} iterations on the
  source positions.  {\it Top:} Position offsets between the first run
  of {\tt fit\_positions} and the second.  Point colors are coded by
  the number of counts (0.35-8.0 keV). {\it Bottom:} Same as {\it
    top}, but comparing our final run of {\tt fit\_positions} to the
  penultimate run.\label{fit_positions}}
\end{figure*}

Once we had measured the source positions using the initial Chandra
astrometric solution that came with the data products, we used the
ACIS-Extract \citep{broos2010} package (version 4994, 2016-09-22) to
measure the positions and photometry at all of these locations in the
data.  We ran ACIS-Extract iteratively, including the task {\tt
  fit\_positions}, to ensure that the software converged on the
position of each source on the ACIS-I detector.  We looked at multiple
possible source positions for each source (``data mean, correlation,
and maximum-likelihood reconstruction'') in each iteration, and we
found that the data mean position both appeared most centered on the
sources and converged reliably for all but the faintest, most off-axis
sources.  We show the impact of our iterative technique in
Figure~\ref{fit_positions}, where we compare the difference in the
data mean positions of the sources between the first two iterations in
one panel and the last two iterations in another panel. Source names
were generated by ACIS-Extract from the input X-ray source positions.

While most sources were observed in only one observation, many were in
overlapping regions, allowing multiple measurements.  In these cases,
the position and position error are from the best (most on-axis)
measurement, as are all of the other values in our final catalog.  

This process resulted in a limiting flux of
3$\times$10$^{-15}$~erg~cm$^{-2}$~s$^{-1}$.

\subsection{Alignment to PHAT}

\begin{figure*}
\begin{center}
\includegraphics[width=5.0in]{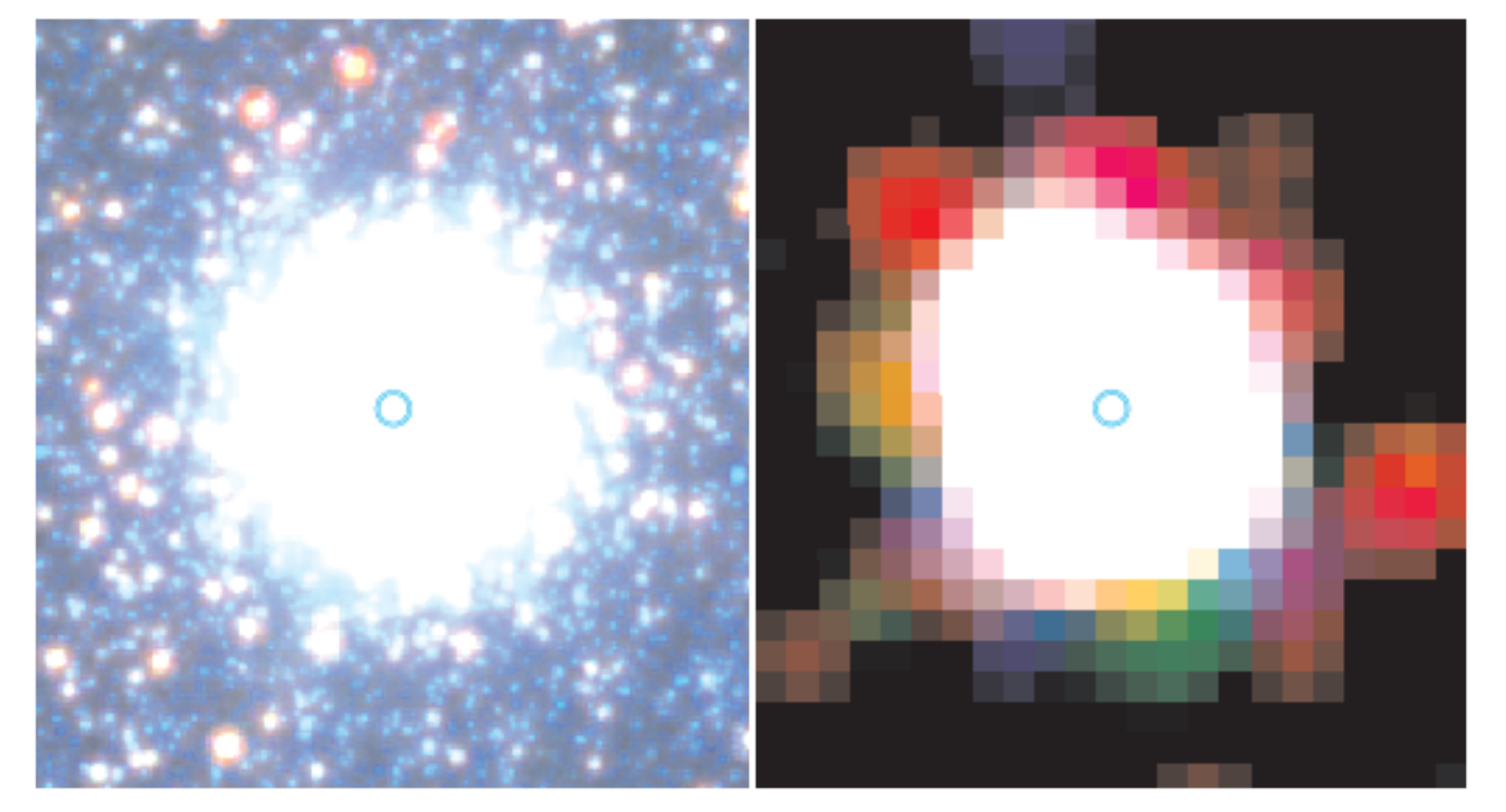}
\includegraphics[width=5.0in]{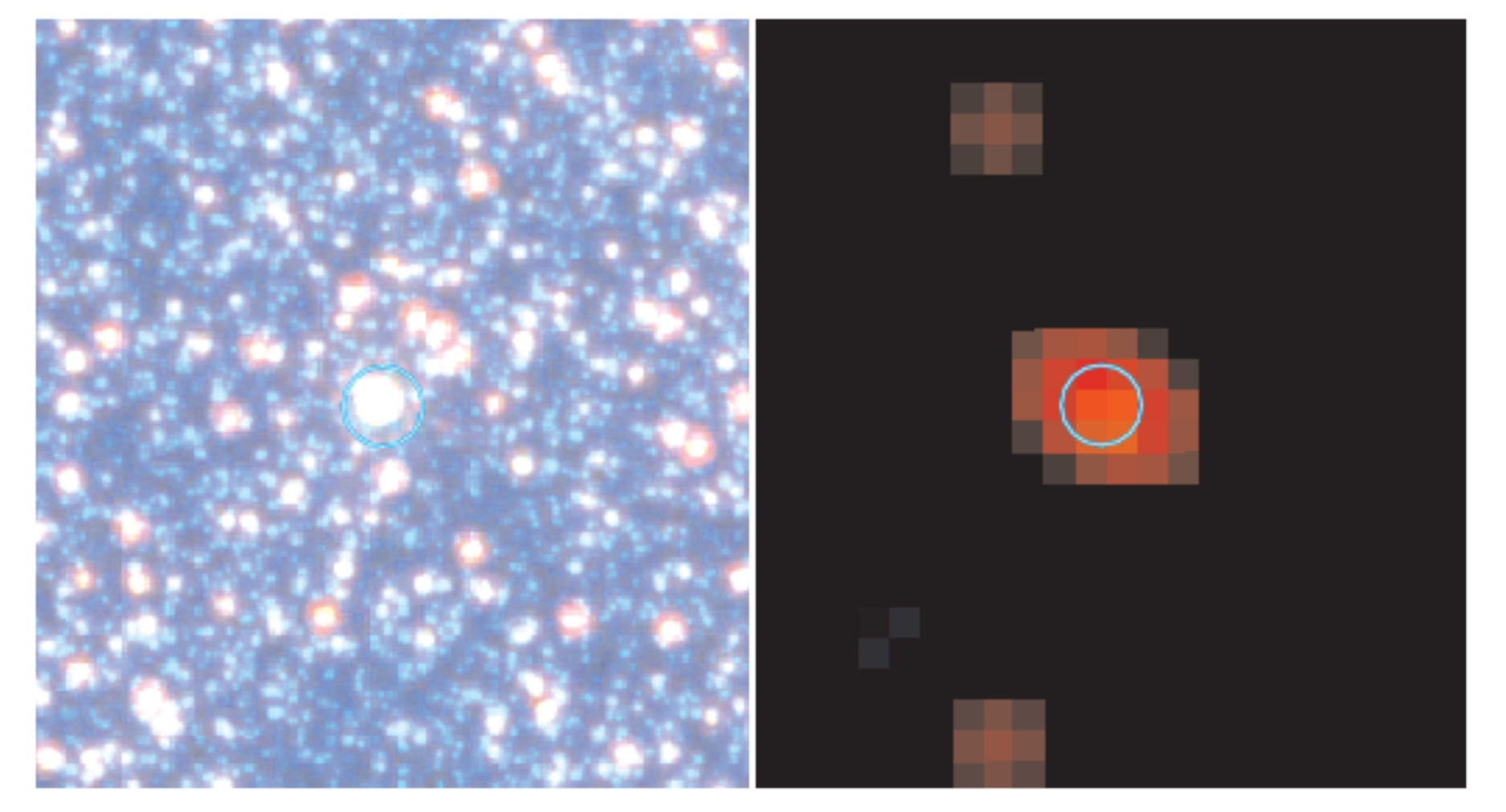}
\end{center}
\caption{Examples of optical sources used to align our {\sl Chandra} imaging to the PHAT imaging shown on 10$''{\times}10''$ PHAT and Chandra images, oriented with north-up and east-left.  {\it Top:} Source CXO~J004429.57+412135.7 in observation 17008 clearly matches a star cluster.  The left panel shows the PHAT image (red=F160W, green=F814W, blue=F475W), with the {\it Chandra} position (blue circle) marked.  The right panel shows the same markings on the color {\it Chandra} image (red=0.35-1 keV, green=1-2 keV, blue=2-8 keV, all smoothed with a Gaussian kernel of radius 3 pixels). {\it Bottom}  Source CXO~J004407.44+412500.0 in observation 17008 is very soft and clearly matches a bright star in the PHAT image.   Because it is so soft, that star is likely to be in the foreground, as bright XRBs typically have hard X-ray spectra. Panels show images and markings corresponding to the same bands and instruments as {\it top}.}
\label{alignment}
\end{figure*}

The PHAT data have astrometric accuracy of $\sim$10 milliarcsec
\citep{williams2014}, while typical raw Chandra data have a 90\%
accuracy of $\sim$800
milliarcsec\footnote{http://cxc.harvard.edu/cal/ASPECT/celmon/}.  To
greatly improve the astrometry of the Chandra data, we aligned the
initial source catalog to the PHAT imaging data using the optimized
Chandra centroids.  We visually inspected all positions on the PHAT
imaging data, making note of which sources corresponded to a clear
foreground star, star cluster, or bright background galaxy.  Examples
of these objects are shown in Figure~\ref{alignment}.  We then used the
IRAF task {\it ccmap} to reset the astrometric solution of the {\it
  Chandra} data to force these positions to align with their
counterparts in the PHAT catalog.  The 26 sources used for alignment
are provided in Table~\ref{aligntab} so that it is clear which sources
were forced to match the PHAT positions.  All Chandra observations had
at least 3 alignment sources, and the corrections were all $<$1$''$.

We fit the positions allowing for rotation, translation, and pixel
scale adjustments.  However, we checked that the solution recovered a
pixel scale of 0.492$''$ (the known Chandra plate scale).  Since there
is little chance that 0.492$''$ in both the X and Y pixel direction
would be the best-fit pixel scale if the matched sources were
incorrect, our recovery of this pixel scale provided confirmation that
the matched sources we used were correctly identified.  For
  background galaxies, we assumed that the X-ray source corresponds to
  the center of the background galaxy.  If the true positions of the
  X-ray sources were far off the galaxy center, the fit to the pixel
  grid would be poor (X and Y plate scales would not likely match).
  Smaller offsets from the center could go unnoticed, but they would
  contribute to the RMS scatter in the fit, which is included in the errors.

The RMS of the astrometric solution returned by {\it ccmap}, suggests
that the {\it Chandra} sources are aligned to the PHAT imaging to a
precision of better than 0.1$''$ (see Table~\ref{obstab}). We have
used the position uncertainty formula of \citet{kim2004} to calculate
the positional uncertainty for the catalog, setting our floor term to
the alignment uncertainty between {\it Chandra} and PHAT for each
observation provided in Table~\ref{obstab}.

\subsection{Counterpart Candidate Identification}\label{sec:colors}

We plotted 1, 2 and 3 $\sigma$ error circles for the Chandra X-ray
sources on the PHAT mosaic images.  These were plotted independently
for each observation so that sources detected in multiple observations
served as a consistency check on our positional alignment to PHAT, and
counterpart candidate.  We examined the X-ray image, color PHAT image,
and UV-only PHAT image at each X-ray source location.  We made a note
of any interesting counterpart candidates, which included bright
stars, blue stars, UV-bright stars, star clusters, or background
galaxies.

\begin{figure*}
\begin{center}
\includegraphics[width=3.5in]{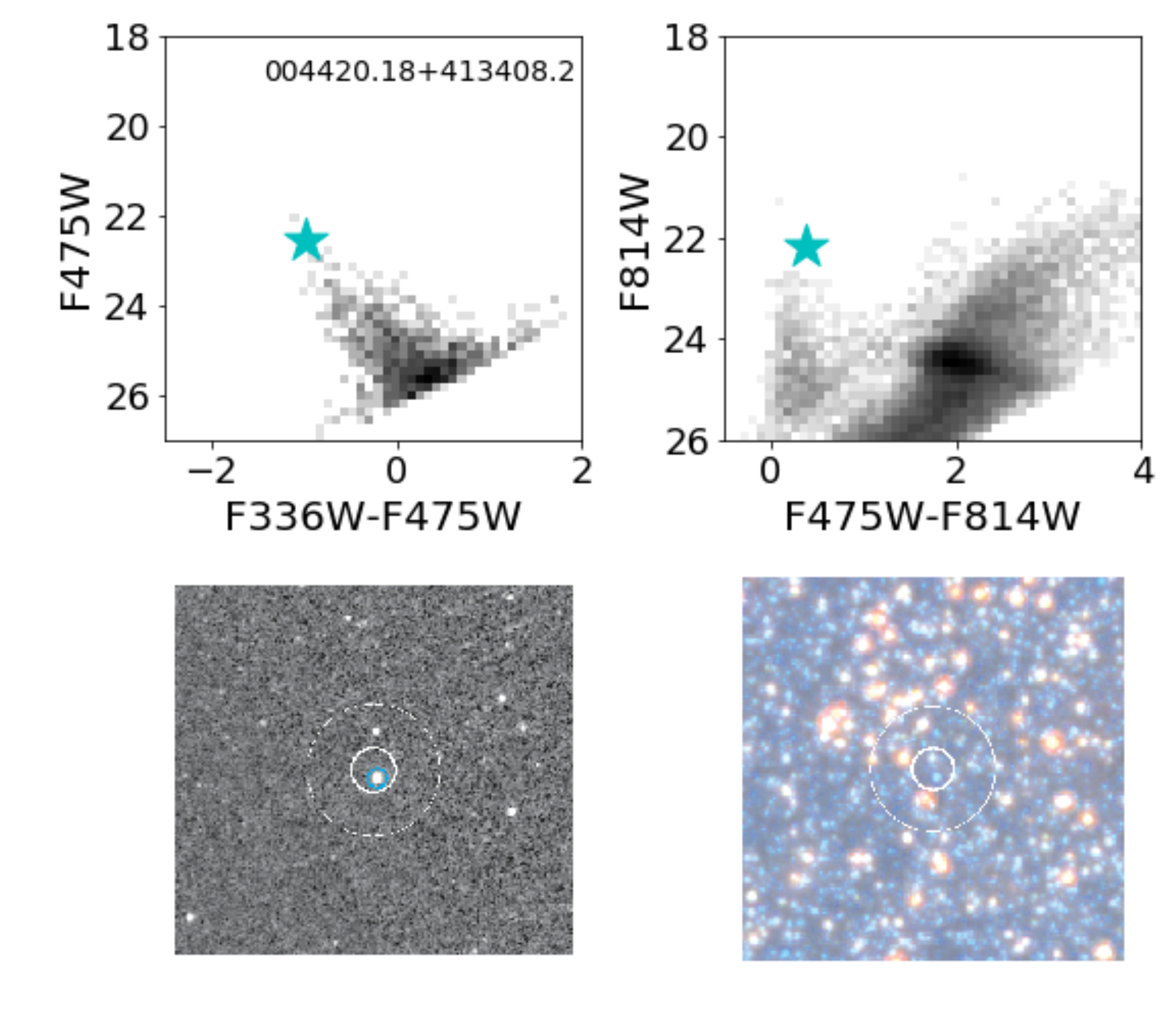}
\includegraphics[width=3.5in]{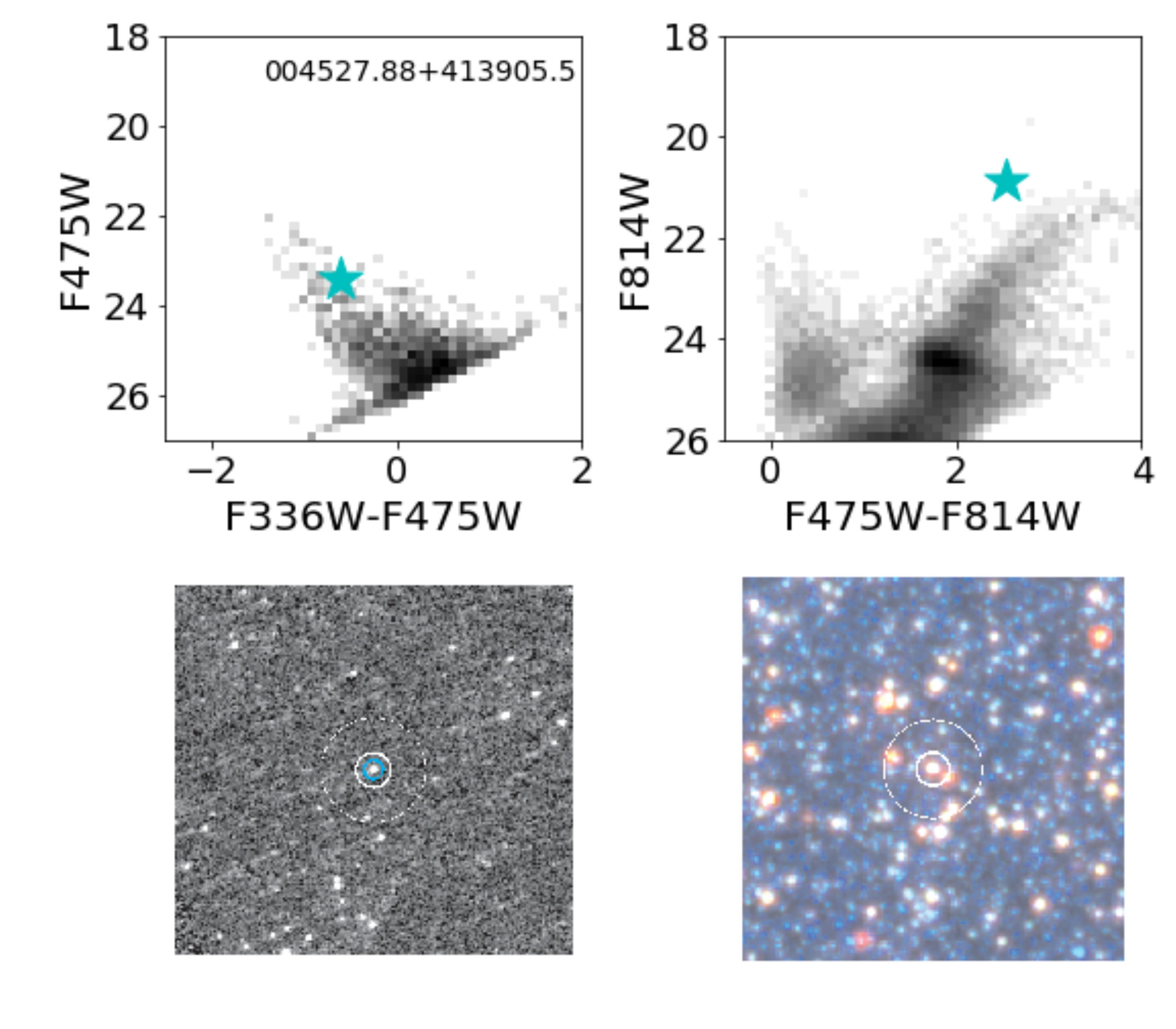}
\includegraphics[width=3.5in]{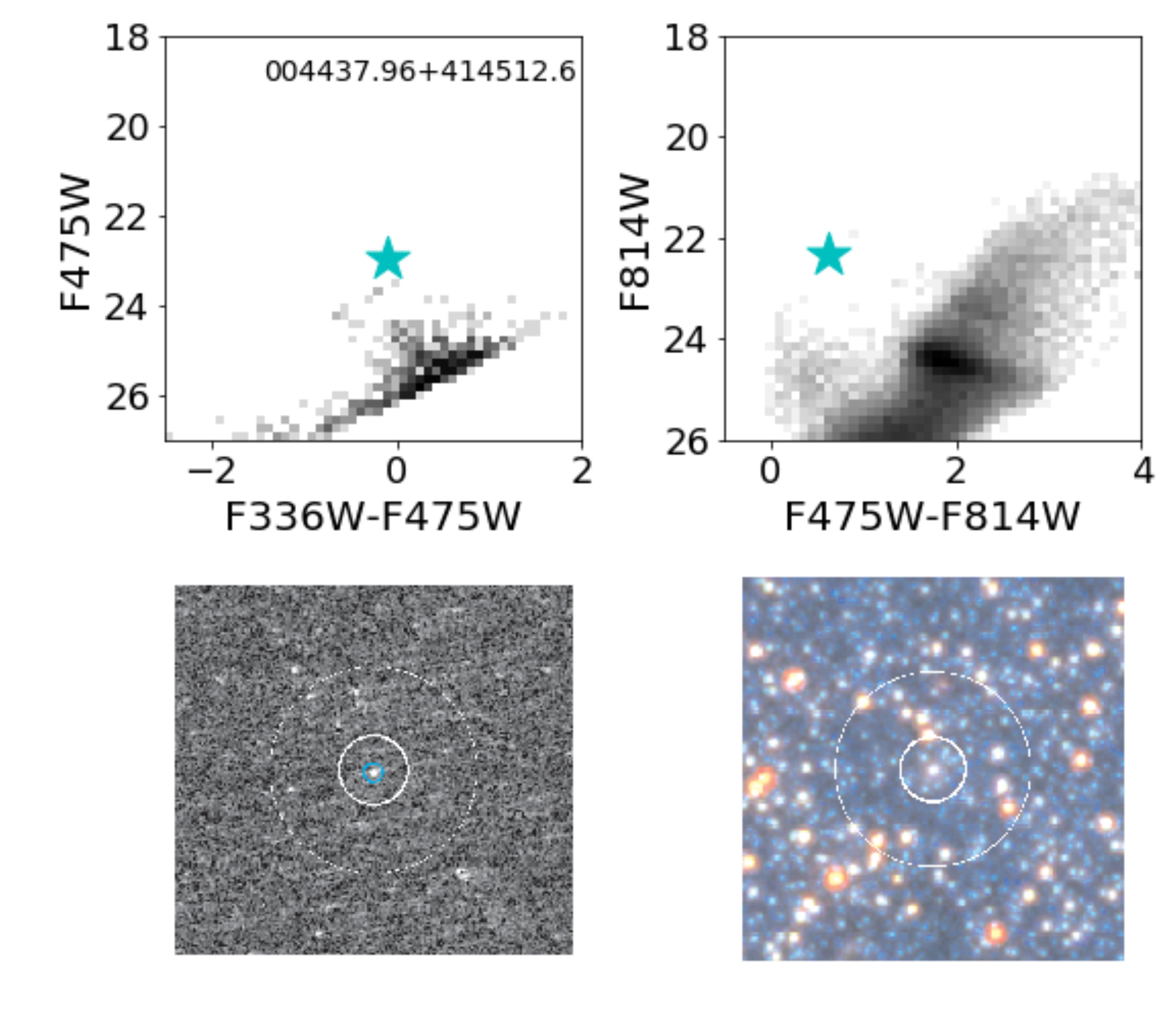}
\end{center}
\caption{Examples of four panel figures showing the identification of
  optical counterpart candidates.  First is a high-quality HMXB
  candidate CXO~J004420.18+413408.2.  Next is a candidate with
  atypical colors, CXO~J004527.88+413905.5.  Finally is a candidate in
  a region without any young stellar population,
  CXO~J004437.96+414512.6. For each source, for example CXO J004527.88+413905.5, the four panels depict {\it Upper-left:} CMD in the near-UV
  (F336W-F475W) from the PHAT catalog of objects within 10$''$ of the
  position of Source CXO~J004527.88+413905.5.  Grayscale shows all
  stars in a 5$''$ circle surrounding the source location for context.
  Purple dots mark other sources within the error circle with measured
  F336W and F475W magnitudes. However, none of these examples contain any such sources. The blue star marks the position of the
  best counterpart candidate as it is UV-bright in addition to being
  spatially coincident with the position of the X-ray source.  {\it
    Upper-right:} Same as Upper-left but for the optical (F475W-F814W)
  CMD.  The object is much brighter and redder in the optical than a
  typical star, suggesting that the SED is being affected by the X-ray
  source (see Section~\ref{sec:colors}).  {\it Lower-left:} F336W PHAT
  image of a 10${\times}$10$''$ region surrounding the X-ray
  position. The 1-$\sigma$ and 3-$\sigma$ error circles are
  overplotted in heavy and light white circles, respectively.  A
  single bright star is circled in blue.  {\it Lower-right:} Color
  PHAT image of the same region (blue is F475W, green is F814W, and
  red is F160W).  In addition to the spatial alignment, this
    object is an excellent counterpart candidate because of its atypical
    colors. \label{finder}}
\end{figure*}

When searching for point source counterparts, we made color-magnitude
diagrams (CMDs) and color-color diagrams of the stars in the PHAT
catalog within 3$\sigma$ of the Chandra source position in
F336W-F475W, F475W-F814W, and F110W-F160W.  We inspected these
diagrams to look for sources with colors that placed them off of the
main CMD features of the survey, or on the upper-main sequence, where
the optical counterparts of HMXBs would tend to reside.  Because HMXBs
have massive secondary stars, if the secondary dominates the optical
light, they would reside on the main sequence in these CMDs; however,
because the primary could have a bright accretion disk that may
irradiate the secondary or produce emission lines, they could be
pushed away from the stellar locus on such CMDs.  Examples of
  these CMDs are shown in Figure~\ref{finder}, where we have plotted
  points for all sources in the error circle with measured F336W and
  F475W magnitudes and marked with a star the counterpart candidate we
  chose by eye.  The color-color diagram of all of our final point
source counterpart candidates is shown in Figure~\ref{colorcolor}.
Figure~\ref{finder} shows and example of each category of point
  source counterpart candidate: those with upper-main sequence colors,
  those with unusual colors, and those not associated with a young
  population.  The unusual colored source has a UV color (upper-left
panel) that is typical of an upper-main sequence star, but its optical
color is quite red, more like a red He-burning star or red supergiant.
Such figures are available for all of the sources in the
  supplemental material, and point source counterpart candidates from
  the PHAT catalog are marked as in the Figure~\ref{finder} example.

The selection of counterpart candidates was iterative.  Two people
searched the images and CMDs independently, compared notes, and
re-examined the positions of sources where there was initial
disagreement until the two lists converged on a set of
candidates. Table~\ref{catalog} gives brief notes on
the sources from this process.  When the process was complete, each
X-ray source either had one optical source that appeared to be a good
candidate (bright star, star cluster, or blue star, UV-bright star, or
background galaxy), or no such candidate.  Objects with no such
candidate could be either highly-extincted (e.g., embedded HMXB, or
AGN behind M31 dust), or too faint in the optical to be detected
(e.g., faint AGN or LMXB).  While our error circles typically contained dozens or more M31 stars, these stars were nearly all similar to the red giant branch stars that are common throughout the galaxy.  We chose only point source candidates that stood out as being brighter and bluer either in the UV or optical than the stars in the surrounding area of the field.  The chances of such an object being within our rather small error circles by chance was relatively low.  Objects of comparable colors and magnitudes have surface densities of $\sim$10$^{-2}$ arcsec$^{-2}$ in PHAT, and our error circles were typically $\lesssim$2 arcsec$^{-2}$, leaving only about a 2\% chance of false positives.

Resolved galaxies are difficult to find in an automated way, since
they are faint and diffuse and are not properly detected or measured
in point source optical catalogs of resolved stars such as that of
\citet{williams2014}.  For example, we cross-correlated our by-eye
background galaxies with those identified by the Andromeda Project
\citep[AP;][]{johnson2015}, which was a crowd-sourcing project in
which the public identified star clusters in the PHAT imaging data.
As a secondary option, users could also mark background galaxies.
They found thousands of relatively bright galaxies in the PHAT
footprint visible in the F475W and F814W bands, 28 of which are in the
1$\sigma$ error circles of X-ray sources in the full sample. However,
we have found that many of the other 79 background galaxies that
coincide with X-ray source positions are very faint and would likely
be missed by those not looking for something specifically at these
locations.  Furthermore, many of these galaxies have most of their
flux in the F160W band, likely because of high absorption in the
optical through the M31 disk. AP limited their search to the F475W and
F814W imaging, and therefore has likely missed many of these very red
background galaxies.

Example images of background galaxy candidates are provided in
Figure~\ref{AGN}, and all such images are included in the supplemental
data.  Figure~\ref{finder} presents an example finding chart from the
PHAT data; the supplemental data include these for all of the sources in the catalog.  These finders allow catalog users to assess
for themselves the veracity of any chosen counterpart candidate.

\subsection{Catalog Cross-Matching}

We cross-correlated all of these sources with the most recent
XMM-Newton catalog (S11) for simple consistency comparisons. This
catalog has a limiting sensitivity of 10$^{-15}$ erg s cm$^{-2}$ in
the 0.2-4.5 keV band.  We matched all sources within 5$''$ of an
XMM-Newton source to allow for the XMM-Newton PSF size. These matches
are all included in our catalog.  There were 311 S11 sources within
our survey area.  We matched 203 of the Chandra sources to 202 of the
S11 sources (S11 1848 matched to both CXO~J004648.19+420855.4 and
CXO~J004648.27+420851.1).  To more directly compare our fluxes with
previous data from XMM-Newton, we used WebPIMMS to determine
conversion factors between ACIS-I count rates in our bands and the S11
0.2-4.5 keV band assuming a power-law spectrum with an index of 1.7
and $N_H$=7$\times$10$^{20}$~cm$^{-2}$.  

A large fraction of fainter the sources in this M31 field vary in
  brightness on long timescales such that they are not detected in all
  observations.  While we found excellent overall agreement for the
203 matched sources, as shown in Figure~\ref{xmm_compare}, the scatter
is sometimes beyond the uncertainties due to intrinsic variability of
sources.  Moreover, most of the fainter sources were only detected in
one set of observations.  There are 170 sources in our catalog that
are not in S11, and 168 of these have 0.35-8 keV fluxes below
1.5$\times$10$^{-14}$ erg cm$^{-2}$ s$^{-1}$
($\sim$10$^{36}$~erg~s$^{-1}$).  Many of these were likely fainter
during the XMM-Newton observations, since the XMM-Newton observations
were sensitive to sources of this brightness.  In the other direction,
there were 109 sources in the S11 catalog in our fields that were not
detected by our observations, and 104 of these had 0.2-4.5 keV fluxes
below 1.5$\times$10$^{-14}$ erg cm$^{-2}$ s$^{-1}$.  The similar
numbers suggest that variability on $\sim$10 year timescales (S11
observations were taken from 2000$-$2008) is the main cause for the
differences.  Such variability at the faint end is similar to that
seen in a dedicated study of the variability of the XLF in NGC~300
\citep{binder2017}, where a large fraction of the sources below
4$\times$10$^{36}$ erg s$^{-1}$ varied significantly between epochs.
X-ray binaries and AGNs are known to vary on such long timescales
  \citep[e.g.,][]{mushotsky1993,mchardy2005,kotze2012}; however, the
  photon statistics on the flux measurements are severely limited for
  these faint sources in M31, making it difficult to quantify their
  variability. These detections are faint, and only put a lower-limit on their
  amplitude.  It is possible that some of these sources could be
  truly transient, and change in brightness by more than a factor of
  100.  More detailed variability analysis of sources in this region
  detected by XMM will be provided in Sasaki et al. (2018, A\&A,
  submitted), as their more sensitive observations allow for more
  precise measurements of the X-ray flux.

Two of the brightest sources not seen by both surveys are
previously-designated transient outbursts.  The brightest source in
S11 that is not in our catalog was S11 1416, which had a 0.2-4.5 keV
flux of 8.55$\times$10$^{-14}$~erg~cm$^{-2}$~s$^{-1}$, and was
  shown by them to be a known transient associated with a nova.  The
brightest source in our catalog that was not in S11
(CXO~J004420.54+413702.3) was designated a transient source by
  Swift \citep{atel2015}.  This source had a 0.2-4.5 keV flux of
3.5$\times$10$^{-13}$~erg~cm$^{-2}$~s$^{-1}$ in our observations.  We
found no clear optical counterpart candidate for this source inside of
its very small error circle, suggesting it is a low-mass X-ray binary.

The only other bright source that does not appear in both catalogs is
CXO~J004427.13+412258.2, which does not appear in any previous X-ray
catalog of M31.  This source has a good point source counterpart
candidate.  In this case, the counterpart candidate is a bright red
star, undetected in the UV.  The PHAT photometry places it above the
tip of the red giant branch, and it is in a region with a high star
formation rate. This could be a potential supergiant X-ray binary, and
would be of particular interest for spectroscopic follow-up.

\section{Results}

Our final source catalog columns are described in Table~\ref{columns},
and the values are provided in Table~\ref{catalog}, including all of
the X-ray measurements for each source found in our 7 observations,
the mean extinction at that location in M31, a one letter code for our
best optical counterpart candidate determination, and descriptions
from the visual inspection of the PHAT images.  We discuss the
characteristics of the sources in detail below.

\subsection{X-ray Properties}

The main goal of this work was to provide exquisite astrometry for the
X-ray sources.  We did not acquire deep enough data for detailed
spectral analysis of the sources.  Much of this work is in a
complementary XMM-Newton program (Sasaki et al. 2018, A\&A,
submitted).  However, we did measure fluxes in many energy bands,
allowing us to examine hardness ratios in the context of the
counterpart candidate types.  In Figure~\ref{HRs}, we plot the
hardness ratios of sources with $>$20 counts (0.35-8 keV,
$\sim$1.2e-14 erg cm$^{-2}$ s$^{-1}$) in our data, color-coding the
points by the candidate type.  We use fluxes in: S=0.35$-$1 keV,
M=1$-$2 keV, and H=2$-$8 keV.  We see the well-known separation of
SNRs and foreground stars which are soft
\citep{pietsch2005,tullmann2011}, congregating at or below
(M-S)/(H+M+S) values of -0.25; however, we also see that all of the
hardest sources ((H-M)/(H+M+S) $>$0.8) with counterpart candidates are
background galaxy candidates. While the uncertainties in these
  ratios, which are${\pm}{\sim}$0.2, make it difficult to reliably
  separate source types in this crowded part of the diagram, this
  distribution hints that sources with very high (H-M)/(H+M+S) ratios
  may be more likely to be background galaxies.  More sensitive observations would be
  necessary to confirm such a possibility.

\subsection{Optical Counterpart Candidates}

We found optical counterpart candidates from the PHAT data for 188
sources.  These include 6 SNRs, 5 star clusters, 12 foreground stars, 107 resolved background galaxies, and 58 point sources.  There were also
185 sources with no clear PHAT counterpart candidate. Below we discuss
each of these source types in turn.

\subsubsection{Supernova Remnants, Star Clusters, and Foreground Stars}

The SNRs and foreground stars are relatively easy to distinguish in
our survey due to their soft X-ray spectra, and their clear detections
in radio \citep[e.g.,][]{braun1990,kong2003,williams2004a,galvin2014} and/or
narrow-band optical wavelengths \citep[e.g.,]{li2014}.

Six previously-known SNRs were detected in our
survey. Table~\ref{tab:snrs} lists the catalog name of the detected
SNRs, the S11 source ID \#, if the remnant was detected in the radio
or the optical, the \citet{li2014} identification number if it exists,
the counts in the 0.35-8.0~keV band, the absorption-corrected luminosity in the 0.35-8.0~keV band, if the source shows extent beyond that expected for
a point source and if there is any evidence of hard emission
(>2.0~keV). The source extraction region created by ACIS-Extract was
examined and adjusted to be larger for these six objects if there was
flux outside of the extraction region.  As SNRs are not point sources,
the measurements in Table~\ref{tab:snrs} are much more reliable for
these sources than those in the point source catalog.

Our survey was conducted with the ACIS-I array to maximize the
field-of-view. Thus the sensitivity to soft sources such as SNRs is
lower than with the S3 CCD on ACIS-S. However, the high angular
resolution of the Chandra data enables a search for emission extended
beyond that expected for a point source and a search for hard emission
that might indicate the presence of a central compact object or pulsar
wind nebula.  Given that the detected counts range from $\sim$5 to 36,
a spectral analysis is not feasible. However, the spatial distribution
of the counts was examined for evidence of extended emission and the
counts above 2.0~keV were examined for evidence of hard emission.

The sources CXO~J004513.88+413615.7, CXO~J004413.49+411954.1, and
CXO~J004451.06+412906.6 show evidence for extended emission, but given
the limited number of counts it is difficult to estimate the size of
the SNR in X-rays.  A deeper observation close to on-axis would be
required to characterize the spatial distribution as ``shell-like'' or
``center-filled''.

Given that the detected counts range from $\sim 14$ to 83 counts, the
luminosities were determined by assuming an {\tt APEC} model in {\tt
  XSPEC} with a temperature of $\mathrm {kT}=0.6~\mathrm{keV}$,
neutral hydrogen column density of
N$_H{=}7.0\times10^{20}~\mathrm{cm}^{-2}$, and solar abundances.  We
fit this model to the data with the only free parameter being the
normalization.  Two of the SNRs, CXO~J004513.88+413615.7 and
CXO~J004451.06+412906.6, appear in more than one observation, such
that the spectral data were combined from the two observations and
weighted response files were created using {\tt specextract} in {\tt CIAO}.  Of the six SNRs detected, only CXO~J004451.06+412906.6 shows
evidence for hard emission, there is a clear excess at high energies
that can not be well fitted by an {\tt APEC} model with
$\mathrm{kT}=0.6~\mathrm{keV}$.  Therefore a power-law component was
added to the spectral model with a fixed index of $\Gamma=2.0$ and a
variable normalization.  The hard counts appear to be centrally
concentrated, whereas the soft counts appear around the periphery of
the hard counts.  A much deeper observation would be required to
confirm this morphology. CXO~J004451.06+412906.6 is a promising
candidate for a central compact object and/or pulsar wind nebula.

The 5 star clusters were well known globular clusters. Two of our sources match to S11 sources that were globular cluster candidates in their survey, but do not appear as clusters in the PHAT data. Source 004343.00+412850.0, which matches to S11 source 1289, had a tentative S11 classification as a globular cluster, but the HST image shows a well-resolved background galaxy at the source location. Source 004353.65+411655.4, which matches to S11 source 1327 had a tentative S11 classification as a globular cluster, but the PHAT data shows no globular cluster at the X-ray source location.  This source is particularly interesting as it is very bright in X-rays but has no outstanding optical source in the error circle.  These traits may make this source an X-ray binary candidate.

The 12 foreground stars we detected have mostly been previously
identified.  These are easily seen from the ground as they appear very
bright in the optical. Thirteen of the objects in our field matched to
S11 are foreground stars. Eleven of the sources classified as
foreground by S11 were independently matched to foreground stars by
us.  In 2 cases, sources that were classified as foreground stars in
S11, were not classified as foreground stars by us.  One of these
(CXO~J004541.06+412752.7) was outside of the PHAT footprint.  The other
(CXO~J004532.11+414527.4) is extended in the PHAT data, suggesting a
background galaxy.  In one case, CXO~J004604.55+414943.7, we see a
foreground star, but S11 classified this object as a SNR candidate.
The source is very far off axis, where the point spread function is
$>$8$''$,  so the position uncertainty is large and we cannot determine
if the source is extended.  The foreground star is just outside of the
one-sigma position uncertainty, so this source classification is still
quite uncertain.

One source was near a foreground star, but the small Chandra
uncertainties show that it is unlikely to be associated.  Source
CXO~J004427.13+412258.2 is just 1$''$ away from of foreground star, but the
star is clearly outside of our Chandra error circle, and the hardness
ratio is harder than those of typical foreground stars. This source is
a newly detected transient by our survey, and the PHAT data show
a bright M31 star in the error circle, making it likely to be an
X-ray binary in M31.

\subsubsection{Background Galaxies}

Figure~\ref{xlfs}, compares the 0.5-7 keV flux distribution of
  our galaxies (90\% of the 0.35-8 keV flux) to the 0.5-7 keV flux
  distribution of the \citet{luo2017} Chandra Deep Field catalog for
  fluxes $>$3$\times$10$^{-15}$ erg cm$^{-2}$ s$^{-1}$.  The
  distribution has been normalized so that both samples have a total
  of 1.  The remarkable similarity suggests that our counterpart
  candidates are correct for these sources, and they are indeed
  background.  Our high fraction of background galaxies (57\% of our counterpart candidates) is
consistent with statistical estimates for the background contamination
in the M31 field.  The 107 background galaxies we have identified can
now be removed from studies attempting to obtain a cleaner sample of
M31 X-ray sources. After removing the 188 sources with counterpart candidates from the sample, there are 185 sources without candidates.  Based on scaling relations \citep[e.g.,][]{lehmer2014}, we expect $\sim$100 LMXBs in
  this region based on the stellar mass. Thus, it is likely that about another
hundred sources in our catalog are still unidentified background
galaxies, which is also consistent with expectations from the Chandra Deep Field, as discussed below.

The by-eye search of the PHAT data identified resolved background
galaxies for roughly half of the total number expected from deep field
statistics. Because of the uncertain impact of the presence of
  M31 on the detection of background sources, we can only make
  sensible estimates of the total number to expect as a sanity check.
Based on the Chandra Deep Field \citep{luo2017}, we expect about half
of our survey area (roughly 0.2 deg$^2$) to be sensitive to background
sources down to 0.35-8 keV fluxes of $\sim$3$\times$10$^{-15}$ erg
cm$^{-2}$ s$^{-1}$.  \citet{luo2017} measured $\sim$1000 sources
deg$^{-2}$ down a 2-7 keV flux level of 10$^{-15}$ erg cm$^{-2}$
s$^{-1}$, suggesting that roughly 200 of the sources we detected are
background galaxies.  If there are another $\sim$90 background
galaxies in our sample ($\sim$200-107), their host galaxies were too
faint or too absorbed to be seen in the PHAT images.  The A$_{V}$
distribution of the source locations (see Section~\ref{av}) does not
appear to have many areas of high extinction, so optical counterpart
intrinsic faintness may be more to blame for non-detections than
absorption.   Some fraction of our point source candidates could also
be AGN where the host galaxy was too faint in the optical for us to
detect. In any case, the number of unidentified sources appears to be consistent with a combination of the expected number of LMXBs given the M31 stellar mass and the expected number of additional background galaxies from the Chandra Deep Field. 

Among the large number of high-quality new galaxy candidates behind
M31 we have found, one of our AGN had been previously identified by
\citet{williams2014hmxbs} as an HMXB candidate.  In that study,
spectra were obtained for blue sources detected in ground based images
within the error circles of the XMM-Newton catalogs of M31.  One of
these was the strong S11 HMXB candidate 1716, which is our source
CXO~J004556.98+414832.0. The blue star was spectroscopically determined to
be a high mass star in M31; however, in this case, the Chandra
position and HST imaging reveals a red background galaxy at the
position of the X-ray source, while the bright blue M31 star is about
2$''$ away from the Chandra position.  Here is a case where the
improved image quality revealed a complex location that was
over-simplified with lower resolution X-ray and ground-based optical
data.

\subsubsection{Point Sources}

In Table~\ref{phat_photometry}, we list the subset of sources with
stellar counterpart candidates in the PHAT survey, along with the PHAT
positions and photometry. Three of our sources
(CXO~J004537.84+414856.7, CXO~J004537.67+415124.4, and
CXO~J004502.33+414943.1) contained both a UV point source and a
background galaxy in the Chandra error circle. In these cases, we made
a note of the galaxy in the notes column; however, we take the UV
emission as the strongest sign of the counterpart.  Thus, these
sources received ``p'' designations in the catalog, and we included
these in our point source analysis.  Figure~\ref{xlfs} shows that the
flux distribution of these sources differs from that of galaxies in
that the point sources tend to be brighter, consistent with this
sample is probing a separate population that contains a larger
fraction of sources in M31.

\begin{figure*}
\begin{center}
\includegraphics[width=6.0in]{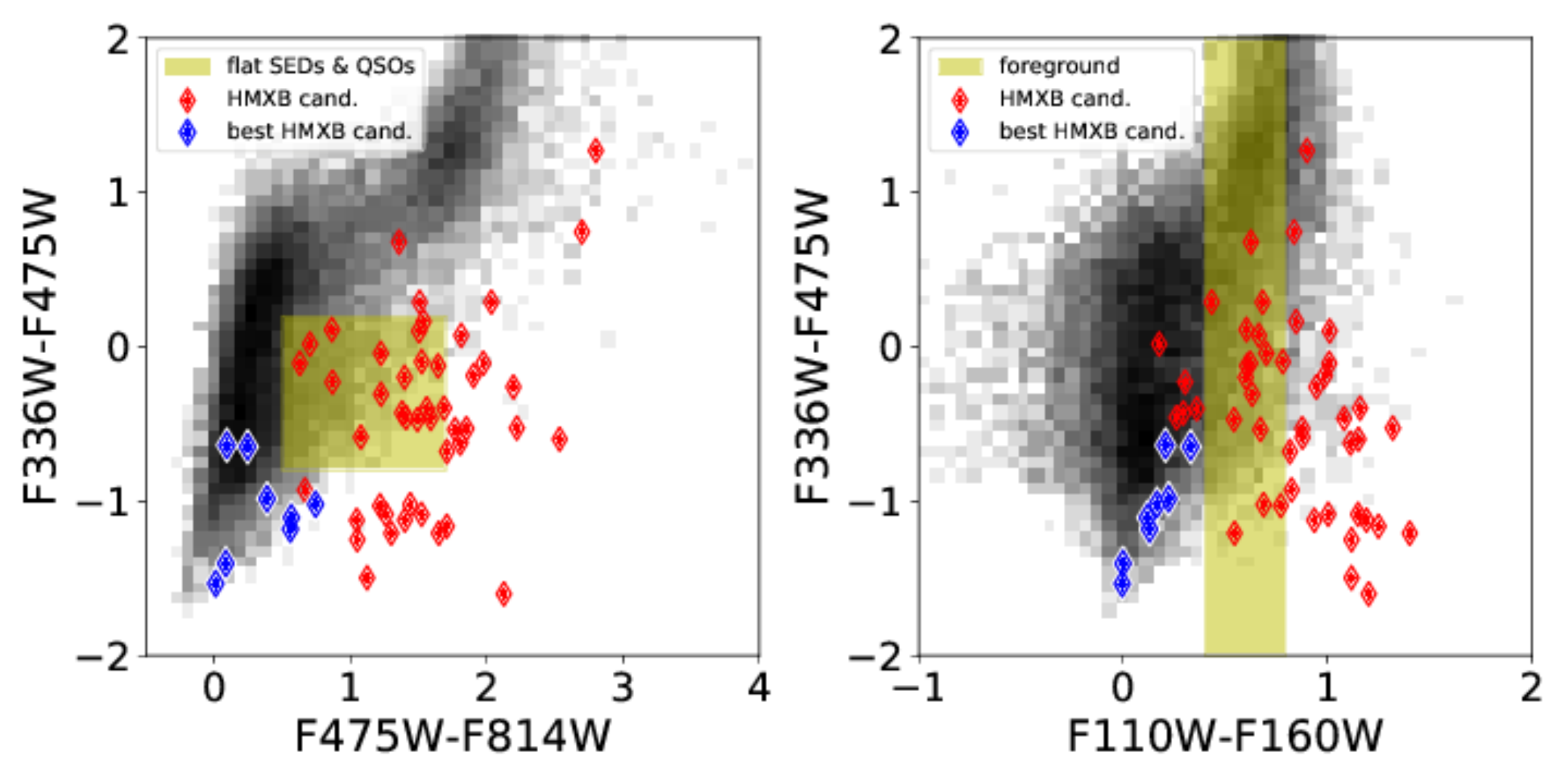}
\end{center}
\caption{UV-optical and UV-IR color-color diagrams of the best stellar optical counterpart candidates.  Other M31 stars near the X-ray sources (black points) follow a clear locus.  The area of the plot where a flat spectrum or a QSO spectrum would fall is shaded yellow in the left panel, and the area where foreground stars would occupy is shaded yellow in the right panel.  Many counterpart candidates fall redward of the stellar locus and in regions expected for sources with non stellar spectra, suggesting that they are not normal single stars, but either binaries in M31 or background galaxies. The counterpart candidates with the bluest colors, typical of young massive stars in M31, are plotted in blue.  These 6 are our best HMXB candidates.}
\label{colorcolor}
\end{figure*}

The PHAT point source counterpart candidates mostly have
  non-standard colors. These colors made them stand out in the PHAT
  imaging allowing us to identify them as very good counterpart
  candidates because any contribution from an accretion disk or
  irradiation from the X-ray source may cause non-standard colors.  As
  shown in Figure~\ref{colorcolor}, some of the colors are consistent
  with the relatively flat spectral energy distributions expected for
  AGN, which argues against these objects belonging to M31.  However,
  since these colors may also be due to a hot X-ray source in a binary
  system and these counterpart candidates are too optically bright to
  be low-mass X-ray binaries, we include all of these sources as
  potential high-mass X-ray binaries.  We plot the X-ray fluxes
  vs. the optical (F475W) magnitudes of these sources in
  Figure~\ref{xlfs}, and the best HMXB candidates generally fall at
  lower X-ray fluxes than the other candidates.

For these PHAT point sources that were not clearly foreground stars
(i.e. saturated sources with bright diffraction spikes), we studied the local stellar populations to shed light on their
  nature. In particular, we would expect HMXBs to reside in young
  regions.  Thus, we used the star formation history results of
\citet{lewis2015} to constrain the age distribution of any co-located
population of stars younger than 80 Myr.  We then take the
  dominant age to be the most likely age of the HMXB system.  Note
  that this is not the time since the binary began producing X-rays,
  but the time since the binary itself was formed. Source
CXO~J004339.06+412117.6 falls in the portion of the PHAT footprint
that was considered too crowed to measure a reliable SFH, leaving 54
sources with local SFH measurements.

Using the local SFHs, we calculate an age probability distribution
function.  We limit the age as older ages can be significantly
contaminated by unassociated stars, which could swamp the signal.  In
addition, neutron star and black hole primaries likely come from
core-collapse SNe, which are produced by stars in this age range.
Thus, the main assumptions are that the source is an HMXB and that
mass transfer onto the compact object began shortly after the compact
object formed.  We show a few examples of these star formation
histories and age distributions in Figure~\ref{age_examples}.

HMXB ages, as inferred from their surrounding populations,
provide sensitive tests of binary formation and evolution models.  The
theoretical work of \citet{belczynski2008} shows the luminosity
distribution to be sensitive to star formation history, and
\citet{linden2010} predict that higher metallicity populations of
bright HMXBs will have a younger distribution than lower metallicity
populations.  These predictions have been qualitatively consistent
with observations; for example, \citet{antoniou2010} and
\citet{williams2013} both find that HMXBs have preferred ages at
$\sim$40-60 Myr, and \citet{antoniou2016} find evidence for a younger
HMXB population (6-25 Myr) in the higher metallicity LMC. In M31, we
probe the highest metallicity \citep[roughly
  solar;][]{venn2000,gregersen2015} extragalactic HMXB sample yet,
putting these predictions to an even stronger observational test.
If HMXB ages are strongly influenced by metallicity then, since
  M31 generally has a higher metallicity than the LMC, we should see
  that in M31, HMXBs are very young (compared to the SMC and LMC).

From these stellar age distributions, we then calculated the stellar
mass in each 0.1 dex wide age bin $<$80 Myr and divided it by the
total amount of stellar mass $<$80 Myr old.  This calculation provides
the fraction of young stars in each bin, which is a proxy for the
probability that the HMXB has that age.  The result is a probability
distribution function for the age of each HMXB candidate living in a
population of stars with ages $<$80 Myr. These probabilities are
provided in Table~\ref{HMXB_ages}. If the counterpart is correct, then
this result is our best estimate of the age of the progenitor of the
compact object, and the X-ray source is likely an HMXB.  In cases
lacking a young population (denoted with a ``c'' in
Table~\ref{phat_photometry}), the source is less likely to be an HMXB;
however, the counterpart candidate is not ruled out, as it may be a
background AGN with a very faint host, a more evolved lower-mass star,
or a runaway massive star.

To investigate the age distribution of the HMXB candidates, we added
the probability distributions together.  Forty of the stellar
candidates have a significant detection of a young ($<$80 Myr)
population in the \citet{lewis2015} maps.  The sum of the
probabilities provides an estimate of the age distribution of our HMXB
candidates.  We show this distribution for the full sample of point
source counterpart candidates, as well as our subsample of the best
HMXB candidates in Figure~\ref{ages}.  In each panel, the histogram
shows the age distribution of the HMXB candidates, and the black lines
show 50 draws of SFHs from random draws from the locations of X-ray
sources associated with background galaxies in our catalog. We do
  not include LMXBs as a separate set of candidates here, as they are
  not distinguishable from undetected background galaxies in our
  data.

All random background galaxy locations show a higher probability of
being older than 30 Myr than being younger.  This result suggests that
these slightly older populations spread over larger areas than the
younger ones.  There is a hint that the full point source counterpart
sample has a higher fraction of areas with $\sim$20 Myr old
populations than the control. If we are more conservative in our
choice of candidates, choosing only candidates that looked
particularly stellar or relatively blue in the PHAT images (marked
with blue diamonds in Figure~\ref{colorcolor}), we find the 8 marked
with an `a' in Table~\ref{phat_photometry}.  Six of these have local
SFH measurements, and all 6 are in regions with recent star formation.
Their age distribution (right panel of Figure~\ref{ages}) has 2 ages:
one at 15-20 Myr, and one at 40-50 Myr.

The 40-50 Myr peak is similar to that found for other samples of HMXB
candidates in nearby galaxies, and may be attributable to the
characteristic timescale for neutron star formation and B-star
activity \citep{antoniou2010,williams2013}.  The 15-20 Myr peak is
observed in the Large Magellanic Cloud \citep{antoniou2016}, and
appears to coincide with a star formation episode 6-25~Myr ago.  These
results could be hinting that the star formation rate in M31 has been
more continuous than in the much smaller SMC and LMC, allowing HMXBs
to form at multiple ages, perhaps through multiple channels.  Seeing
both ages in M31 suggests that both formation timescales occur at high
metallicity, which would point to differences between the Magellanic Cloud populations being more attributable to star formation history than to metallicity. 


\subsection{Extinction}\label{av}

In our catalog, 341 of the sources are located inside the area of the
PHAT footprint covered by the \citet{dalcanton2015} extinction map.
For all of these locations, we found the $A_{\rm V}$ of the 3.3$''$
pixel on the \citet{dalcanton2015} map covering the location. This
value is the mean extinction of the M31 stars in that
3.3$''{\times}3.3''$ region. We report this value in our catalog.
However, we note that any individual source may not actually
experience the mean extinction depending on its position along the
line of sight in the M31 disk.  We note that Sasaki et al. (2018,
A\&A, submitted) has performed spectral fits to the XMM data of many
of these sources, finding higher $N_H$ values for the background
galaxies, confirming that the mean extinction in M31 does not reflect
the full absorption column toward background objects.

We plot histograms of these extinction values in Figure~\ref{avhist}
for the complete $A_{\rm V}$ map, the total sample, and for subsamples
broken down by counterpart candidate.  Overall, the distributions do
not show strong differences. There is no evidence that the X-ray
sources or their counterpart candidates are preferentially found in
regions of low extinction, as might be expected if dust were strongly
inhibiting our ability to detect sources or counterpart candidates.
Several of the sources in the highest extinction locations are
associated with resolved background galaxies, but the background
galaxy distribution has similar extinction values as the full sample
distribution. 

These distributions all suggest that counterpart candidate
determination is not strongly correlated with extinction in M31.
Thus, extinction does not appear to be a limiting factor in finding
counterpart candidates for X-ray sources in M31.  The displaced
  colors of many of the sources shown in Figure~\ref{colorcolor} do not
  appear to be due to dust, which is not surprising since dust would
  make them much redder than observed in F336W-F475W.  Apparently,
intrinsic faintness of the counterparts, whether background galaxies
or M31 stars, is limiting our ability to find more counterpart
candidates.  We note that we do not find any good stellar or star
cluster counterpart candidates in regions of M31 with mean $A_{\rm
  V}{>}2.1$, but there are very few X-ray sources detected in these
dusty regions, and perhaps the X-ray sources in such regions are
continuing to go undetected. Such regions are also very rare in M31,
which is consistent with the ease of identifying background galaxies
through modest dust columns.

\section{Conclusions}

We have obtained {\sl Chandra} imaging covering a large fraction of
M31 with {\sl Hubble Space Telescope} imaging obtained by the
Panchromatic Hubble Andromeda Treasury (PHAT) survey.  Combining these
data sets, we have produced a catalog of X-ray sources along with
their most likely optical counterparts from HST. These optical
  counterpart candidate identifications allow background galaxies and
high-mass X-ray binaries to be separated from other potential hard
sources, such as low-mass X-ray binaries.

\begin{itemize}
\item We find that most counterpart candidates are resolved background galaxies, and that there are over 100 of these, which is consistent with the majority of X-ray sources in the M31 disk field being background contaminants if we assume a similar fraction of the sources with no counterpart candidate are undetected fainter background galaxies.  This assumption is consistent both with the expected number of background galaxies estimated from the Chandra Deep Field and with the expected number of LMXBs estimated from the PHAT stellar mass.
\item We find about a third of the point source counterpart candidates are not associated with any young stellar populations.
\item The number of optical point source candidates (58) is larger than the expected number of bright HMXBs in this region, but it is similar to expectations if about half of the 40 candidates in regions with young stellar populations are indeed HMXBs.
\item We find 8 of the point source counterpart candidates have colors typical of single stars, suggesting that many of the point sources in this sample are background galaxies. The number of good HMXB candidates is somewhat below the number expected from the star formation rate and number of OB stars in the region surveyed; however, some of the other point source candidates could be HMXBs with odd colors due to binarity.  Further observations will be necessary to determine if M31 actually has as many bright HMXBs as predicted by scaling relations.
\item We find that the age distribution of the young populations
  surrounding the point source counterpart candidates (including the 8
  with typical star colors), is peaked at 15-20 Myr and 40-50 Myr in
  agreement with previous studies in other nearby galaxies \citep{antoniou2010,williams2013,antoniou2016}, but at higher metallicity.
\item Based on the extinction results here, dust does not appear to be
  significantly impeding the searches for optical counterparts.

\end{itemize}

The production of this catalog is only the beginning of the
Chandra-PHAT program.  We are currently working to perform and study
MCMC fits to the spectral energy distribution from the PHAT photometry
of all of the sources within 3-$\sigma$ of X-ray sources using the
Bayesian Extinction and Stellar Tool
\citep[BEAST][]{gordon2016} similar to those done for the NuSTAR sources in \citet{lazzarini2018}. These fits will likely provide further
confirmation of our original classifications presented here, and may
result in a few new classifications in cases where there were no
obvious candidates in our CMDs or images.  Furthermore, the SED fits
should allow us to provide physical parameters for the secondaries.
Finally, follow-up spectroscopy of our optical point source candidates
will help identify more HMXBs, type their secondaries, and measure
orbital periods.  In turn, we can use all of these measurements in
addition to the local star formation histories to place new
constraints on X-ray binary formation and evolution models by
improving the statistics on their age and mass distributions which the
models should reproduce.

Support for this work was provided by Chandra Award Number GO5-16085X
issued by the Chandra X-ray Observatory Center, which is operated by
the Smithsonian Astrophysical Observatory for and on behalf of the
National Aeronautics and Space Administration under contract
NAS8-03060.  M.S.\ acknowledges support by the Deutsche
Forschungsgemeinschaft (DFG) through the Heisenberg fellowship SA
2131/3-1 and the Heisenberg professor grant SA 2131/5-1. B.W., M.S. \&
P.P. acknowledge partial support for this research through the Chandra
Research Visitors Program.


\begin{thebibliography}{}

\bibitem[\protect\citeauthoryear{{Antoniou} \& {Zezas}}{{Antoniou} \&
    {Zezas}}{2016}]{antoniou2016} {Antoniou}, V., \& {Zezas}, A. 2016,
  \mnras, 459, 528

\bibitem[\protect\citeauthoryear{{Antoniou} et~al.}{{Antoniou}
  et~al.}{2010}]{antoniou2010}
{Antoniou}, V., {Zezas}, A., {Hatzidimitriou}, D.,  \& {Kalogera}, V. 2010,
  \apjl, 716, L140

\bibitem[\protect\citeauthoryear{{Belczynski} et~al.}{{Belczynski}
  et~al.}{2008}]{belczynski2008}
{Belczynski}, K., {Kalogera}, V., {Rasio}, F.~A., {Taam}, R.~E., {Zezas}, A.,
  {Bulik}, T., {Maccarone}, T.~J.,  \& {Ivanova}, N. 2008, \apjs, 174, 223

\bibitem[\protect\citeauthoryear{{Binder} et~al.}{{Binder}
  et~al.}{2017}]{binder2017}
{Binder}, B., {Gross}, J., {Williams}, B.~F., {Eracleous}, M., {Gaetz}, T.~J.,
  {Plucinsky}, P.~P.,  \& {Skillman}, E.~D. 2017, \apj, 834, 128

\bibitem[\protect\citeauthoryear{{Binder} et~al.}{{Binder}
  et~al.}{2013}]{binder2013}
{Binder}, B., {Williams}, B.~F., {Eracleous}, M., {Gaetz}, T.~J., {Kong},
  A.~K.~H., {Skillman}, E.~D.,  \& {Weisz}, D.~R. 2013, \apj, 763, 128

\bibitem[\protect\citeauthoryear{{Binder} et~al.}{{Binder}
  et~al.}{2012}]{binder2012}
{Binder}, B., et~al. 2012, \apj, 758, 15

\bibitem[\protect\citeauthoryear{{Binder} et~al.}{{Binder}
  et~al.}{2015}]{binder2015}
{Binder}, B., et~al. 2015, \aj, 150, 94

\bibitem[\protect\citeauthoryear{{Braun}}{{Braun}}{1990}]{braun1990}
{Braun}, R. 1990, \apjs, 72, 761

\bibitem[\protect\citeauthoryear{{Broos} et~al.}{{Broos}
  et~al.}{2010}]{broos2010}
{Broos}, P.~S., {Townsley}, L.~K., {Feigelson}, E.~D., {Getman}, K.~V.,
  {Bauer}, F.~E.,  \& {Garmire}, G.~P. 2010, \apj, 714, 1582

\bibitem[\protect\citeauthoryear{{Dalcanton} et~al.}{{Dalcanton}
  et~al.}{2015}]{dalcanton2015}
{Dalcanton}, J.~J., et~al. 2015, \apj, 814, 3

\bibitem[\protect\citeauthoryear{{Dalcanton} et~al.}{{Dalcanton}
  et~al.}{2012}]{dalcanton2012}
{Dalcanton}, J.~J., et~al. 2012, \apjs, 200, 18

\bibitem[\protect\citeauthoryear{{Dalcanton} et~al.}{{Dalcanton}
  et~al.}{2009}]{dalcanton2009}
{Dalcanton}, J.~J., et~al. 2009, \apjs, 183, 67

\bibitem[\protect\citeauthoryear{{Di Stefano} et~al.}{{Di Stefano}
  et~al.}{2004}]{distefano2004}
{Di Stefano}, R., et~al. 2004, \apj, 610, 247

\bibitem[Fruscione et al.(2006)]{ciao} Fruscione, A., et al.\ 2006, \procspie, 6270

\bibitem[\protect\citeauthoryear{{Galvin} \& {Filipovic}}{{Galvin} \&
  {Filipovic}}{2014}]{galvin2014}
{Galvin}, T.~J.,  \& {Filipovic}, M.~D. 2014, Serbian Astronomical Journal,
  189, 15

\bibitem[\protect\citeauthoryear{{Garcia} et~al.}{{Garcia}
  et~al.}{2010}]{garcia2010}
{Garcia}, M.~R., et~al. 2010, \apj, 710, 755

\bibitem[\protect\citeauthoryear{{Gordon} et~al.}{{Gordon}
  et~al.}{2016}]{gordon2016}
{Gordon}, K.~D., et~al. 2016, \apj, 826, 104

\bibitem[\protect\citeauthoryear{{Gregersen} et~al.}{{Gregersen}
  et~al.}{2015}]{gregersen2015}
{Gregersen}, D., et~al. 2015, \aj, 150, 189

\bibitem[\protect\citeauthoryear{{Grimm}, {Gilfanov}, \& {Sunyaev}}{{Grimm}
  et~al.}{2003}]{grimm2003}
{Grimm}, H.-J., {Gilfanov}, M.,  \& {Sunyaev}, R. 2003, \mnras, 339, 793

\bibitem[\protect\citeauthoryear{{Hatzidimitriou} et~al.}{{Hatzidimitriou}
  et~al.}{2006}]{hatzidimitriou2006}
{Hatzidimitriou}, D., {Pietsch}, W., {Misanovic}, Z., {Reig}, P.,  \& {Haberl},
  F. 2006, \aap, 451, 835

\bibitem[\protect\citeauthoryear{{Henze} et~al.}{{Henze}
  et~al.}{2015}]{atel2015}
{Henze}, M., {Sasaki}, M., {Haberl}, F.,  \& {Hatzidimitriou}, D. 2015, The
  Astronomer's Telegram, 8227

\bibitem[\protect\citeauthoryear{{Johnson} et~al.}{{Johnson}
  et~al.}{2015}]{johnson2015}
{Johnson}, L.~C., et~al. 2015, \apj, 802, 127

\bibitem[\protect\citeauthoryear{{Kaaret}}{{Kaaret}}{2002}]{kaaret2002}
{Kaaret}, P. 2002, \apj, 578, 114

\bibitem[\protect\citeauthoryear{{Kim} et~al.}{{Kim} et~al.}{2004}]{kim2004}
{Kim}, D.-W., et~al. 2004, \apjs, 150, 19

\bibitem[\protect\citeauthoryear{{Kong} et~al.}{{Kong} et~al.}{2002}]{kong2002}
{Kong}, A.~K.~H., {Garcia}, M.~R., {Primini}, F.~A., {Murray}, S.~S., {Di
  Stefano}, R.,  \& {McClintock}, J.~E. 2002, \apj, 577, 738

\bibitem[\protect\citeauthoryear{{Kong} et~al.}{{Kong} et~al.}{2003}]{kong2003}
{Kong}, A.~K.~H., {Sjouwerman}, L.~O., {Williams}, B.~F., {Garcia}, M.~R.,  \&
  {Dickel}, J.~R. 2003, \apjl, 590, L21

\bibitem[Kotze \& Charles(2012)]{kotze2012} Kotze, M.~M., \& Charles, P.~A.\ 2012, \mnras, 420, 1575 

\bibitem[Lazzarini et al.(2018)]{lazzarini2018} Lazzarini, M., Hornschemeier, A.~E., Williams, B.~F., et al.\ 2018, arXiv:1806.03305 

\bibitem[\protect\citeauthoryear{{Lee} \& {Lee}}{{Lee} \& {Lee}}{2014}]{li2014}
{Lee}, J.~H.,  \& {Lee}, M.~G. 2014, \apj, 786, 130

\bibitem[Lehmer et al.(2014)]{lehmer2014} Lehmer, B.~D., Berkeley, M., Zezas, A., et al.\ 2014, \apj, 789, 52 



\bibitem[\protect\citeauthoryear{{Lewis} et~al.}{{Lewis}
  et~al.}{2015}]{lewis2015}
{Lewis}, A.~R., et~al. 2015, \apj, 805, 183

\bibitem[\protect\citeauthoryear{{Li} et~al.}{{Li} et~al.}{2011}]{li2011}
{Li}, Z., {Garcia}, M.~R., {Forman}, W.~R., {Jones}, C., {Kraft}, R.~P., {Lal},
  D.~V., {Murray}, S.~S.,  \& {Wang}, Q.~D. 2011, \apjl, 728, L10

\bibitem[\protect\citeauthoryear{{Linden} et~al.}{{Linden}
  et~al.}{2010}]{linden2010}
{Linden}, T., {Kalogera}, V., {Sepinsky}, J.~F., {Prestwich}, A., {Zezas}, A.,
  \& {Gallagher}, J.~S. 2010, \apj, 725, 1984

\bibitem[\protect\citeauthoryear{{Luo} et~al.}{{Luo} et~al.}{2017}]{luo2017}
{Luo}, B., et~al. 2017, \apjs, 228, 2

\bibitem[\protect\citeauthoryear{{McConnachie} et~al.}{{McConnachie}
  et~al.}{2005}]{mcconnachie2005}
{McConnachie}, A.~W., {Irwin}, M.~J., {Ferguson}, A.~M.~N., {Ibata}, R.~A.,
  {Lewis}, G.~F.,  \& {Tanvir}, N. 2005, \mnras, 356, 979

\bibitem[McHardy et al.(2005)]{mchardy2005} McHardy, I.~M., Gunn, K.~F., Uttley, P., \& Goad, M.~R.\ 2005, \mnras, 359, 1469 

\bibitem[Mushotzky et al.(1993)]{mushotsky1993} Mushotzky, R.~F., Done, C., \& Pounds, K.~A.\ 1993, \araa, 31, 717 


\bibitem[\protect\citeauthoryear{{Pietsch}, {Freyberg}, \& {Haberl}}{{Pietsch}
  et~al.}{2005}]{pietsch2005}
{Pietsch}, W., {Freyberg}, M.,  \& {Haberl}, F. 2005, \aap, 434, 483

\bibitem[\protect\citeauthoryear{{Shaw Greening} et~al.}{{Shaw Greening}
  et~al.}{2009}]{ShawGreening2009}
{Shaw Greening}, L., {Barnard}, R., {Kolb}, U., {Tonkin}, C.,  \& {Osborne},
  J.~P. 2009, \aap, 495, 733

\bibitem[\protect\citeauthoryear{{Stiele} et~al.}{{Stiele}
  et~al.}{2011}]{stiele2011}
{Stiele}, H., {Pietsch}, W., {Haberl}, F., {Hatzidimitriou}, D., {Barnard}, R.,
  {Williams}, B.~F., {Kong}, A.~K.~H.,  \& {Kolb}, U. 2011, \aap, 534, A55

\bibitem[\protect\citeauthoryear{{Supper} et~al.}{{Supper}
  et~al.}{2001}]{supper2001}
{Supper}, R., {Hasinger}, G., {Lewin}, W.~H.~G., {Magnier}, E.~A., {van
  Paradijs}, J., {Pietsch}, W., {Read}, A.~M.,  \& {Tr{\"u}mper}, J. 2001,
  \aap, 373, 63

\bibitem[\protect\citeauthoryear{{Supper} et~al.}{{Supper}
  et~al.}{1997}]{supper1997}
{Supper}, R., {Hasinger}, G., {Pietsch}, W., {Truemper}, J., {Jain}, A.,
  {Magnier}, E.~A., {Lewin}, W.~H.~G.,  \& {van Paradijs}, J. 1997, \aap, 317,
  328

\bibitem[\protect\citeauthoryear{{T{\"u}llmann} et~al.}{{T{\"u}llmann}
  et~al.}{2011}]{tullmann2011}
{T{\"u}llmann}, R., et~al. 2011, \apjs, 193, 31

\bibitem[\protect\citeauthoryear{{van Speybroeck} et~al.}{{van Speybroeck}
  et~al.}{1979}]{vanspeybroeck1979}
{van Speybroeck}, L., {Epstein}, A., {Forman}, W., {Giacconi}, R., {Jones}, C.,
  {Liller}, W.,  \& {Smarr}, L. 1979, \apjl, 234, L45

\bibitem[\protect\citeauthoryear{{Venn} et~al.}{{Venn} et~al.}{2000}]{venn2000}
{Venn}, K.~A., {McCarthy}, J.~K., {Lennon}, D.~J., {Przybilla}, N.,
  {Kudritzki}, R.~P.,  \& {Lemke}, M. 2000, \apj, 541, 610

\bibitem[\protect\citeauthoryear{{Vulic}, {Gallagher}, \& {Barmby}}{{Vulic}
  et~al.}{2016}]{vulic2016}
{Vulic}, N., {Gallagher}, S.~C.,  \& {Barmby}, P. 2016, \mnras, 461, 3443

\bibitem[\protect\citeauthoryear{{Williams}}{{Williams}}{2003}]{williams2003}
{Williams}, B.~F. 2003, \aj, 126, 1312

\bibitem[\protect\citeauthoryear{{Williams} et~al.}{{Williams}
  et~al.}{2013}]{williams2013}
{Williams}, B.~F., {Binder}, B.~A., {Dalcanton}, J.~J., {Eracleous}, M.,  \&
  {Dolphin}, A. 2013, \apj, 772, 12

\bibitem[Williams et al.(2017)]{williams2017} Williams, B.~F., Dolphin, A.~E., Dalcanton, J.~J., et al.\ 2017, \apj, 846, 145 

\bibitem[\protect\citeauthoryear{{Williams} et~al.}{{Williams}
  et~al.}{2004a}]{williams2004}
{Williams}, B.~F., {Garcia}, M.~R., {Kong}, A.~K.~H., {Primini}, F.~A., {King},
  A.~R., {Di Stefano}, R.,  \& {Murray}, S.~S. 2004a, \apj, 609, 735

\bibitem[\protect\citeauthoryear{{Williams} et~al.}{{Williams}
  et~al.}{2005a}]{williams2005c}
{Williams}, B.~F., {Garcia}, M.~R., {McClintock}, J.~E., {Kong}, A.~K.~H.,
  {Primini}, F.~A.,  \& {Murray}, S.~S. 2005a, \apj, 628, 382

\bibitem[\protect\citeauthoryear{{Williams} et~al.}{{Williams}
  et~al.}{2005b}]{williams2005a}
{Williams}, B.~F., {Garcia}, M.~R., {McClintock}, J.~E., {Primini}, F.~A.,  \&
  {Murray}, S.~S. 2005b, \apj, 632, 1086

\bibitem[\protect\citeauthoryear{{Williams} et~al.}{{Williams}
  et~al.}{2005c}]{williams2005b}
{Williams}, B.~F., {Garcia}, M.~R., {Primini}, F.~A., {McClintock}, J.~E.,  \&
  {Murray}, S.~S. 2005c, \apj, 631, 832

\bibitem[\protect\citeauthoryear{{Williams} et~al.}{{Williams}
  et~al.}{2014a}]{williams2014hmxbs}
{Williams}, B.~F., et~al. 2014a, \mnras, 443, 2499

\bibitem[\protect\citeauthoryear{{Williams} et~al.}{{Williams}
  et~al.}{2014b}]{williams2014}
{Williams}, B.~F., et~al. 2014b, \apjs, 215, 9

\bibitem[\protect\citeauthoryear{{Williams} et~al.}{{Williams}
  et~al.}{2004b}]{williams2004a}
{Williams}, B.~F., {Sjouwerman}, L.~O., {Kong}, A.~K.~H., {Gelfand}, J.~D.,
  {Garcia}, M.~R.,  \& {Murray}, S.~S. 2004b, \apj, 615, 720

\end{thebibliography}

\clearpage


\begin{deluxetable}{ccccccc}
\tablecaption{Chandra-PHAT Observations}
\tabletypesize{\scriptsize}
\tablehead{
\colhead{ObsID} &
\colhead{RA\_NOM (deg)} &
\colhead{DEC\_NOM (deg)} &
\colhead{ROLL\_NOM (deg)} &
\colhead{Date (YYYY-MM-DD)} &
\colhead{Exptime (s)} & 
\colhead{RMS ($''$)}
}
\startdata
17008 & 11.06541197 & 41.38758162 & 181.58 & 2015-10-06 & 49141 & 0.08\\
17009 & 11.01739502 & 41.57751223 & 215.29 & 2015-10-26 & 49405 & 0.08\\
17010 & 11.24607031 & 41.53432975 & 202.70 & 2015-10-19 & 49423 & 0.07\\
17011 & 11.37564751 & 41.72352482 & 184.29 & 2015-10-08 & 49429 & 0.07\\
17012 & 11.19575165 & 41.86086452 & 196.22 & 2015-10-11 & 48440 & 0.05\\
17013 & 11.53451858 & 41.95795668 & 199.11 & 2015-10-17 & 44790 & 0.08\\
17014 & 11.58803371 & 42.15469119 & 180.22 & 2015-10-09 & 49139 & 0.08
\enddata
\label{obstab}
\end{deluxetable}

\begin{deluxetable}{ccccc}
\tablecaption{Chandra-PHAT Alignment Sources}
\tablehead{
\colhead{ObsID} &
\colhead{Catalog Name (CXO~J)} &
\colhead{PHAT RA} &
\colhead{PHAT Dec} &
\colhead{Description}
}
17009 & 004345.50+413657.5 & 10.9395926 & 41.6160038 & Globular Cluster\\
17009 & 004350.29+413248.8 & 10.9595925 & 41.5468718 & Background galaxy\\
17008 & 004356.43+412203.0 & 10.9851124 & 41.3674795 & Globular Cluster\\
17008 & 004407.44+412500.0 & 11.0310358 & 41.4166779 & Bright Star\\
17009 & 004425.57+413633.6 & 11.1065203 & 41.6093377 & Foreground Star\\
17008 & 004429.57+412135.7 & 11.1231821 & 41.3599211 & Globular Cluster\\
17012 & 004442.71+415340.8 & 11.1779661 & 41.8946660 & Background galaxy\\
17012 & 004443.43+415231.1 & 11.1809490 & 41.8753160 & Background galaxy\\
17012 & 004444.88+415154.0 & 11.1870083 & 41.8650016 & Background galaxy\\
17011 & 004525.63+414315.4 & 11.3567486 & 41.7209178 & Background galaxy\\
17010 & 004525.67+413158.2 & 11.3569586 & 41.5328505 & Background galaxy\\
17010 & 004526.81+413217.4 & 11.3616871 & 41.5381725 & Background galaxy\\
17010 & 004527.30+413254.1 & 11.3637671 & 41.5483491 & Background galaxy\\
17010 & 004528.24+412943.9 & 11.3676874 & 41.4955348 & Background galaxy\\
17011 & 004533.35+414330.5 & 11.3889692 & 41.7251433 & Background galaxy\\
17011 & 004545.57+413942.1 & 11.4398766 & 41.6617307 & Globular Cluster\\
17011 & 004556.98+414832.0 & 11.4874508 & 41.8088835 & Background galaxy\\
17013 & 004559.39+415835.9 & 11.4974523 & 41.9766692 & Background galaxy\\
17014 & 004608.29+421054.6 & 11.5345977 & 42.1818576 & Background galaxy\\
17013 & 004609.61+415440.1 & 11.5400766 & 41.9111281 & Background galaxy\\
17014 & 004625.31+420938.0 & 11.6054444 & 42.1605639 & Background galaxy\\
17013 & 004625.71+415526.2 & 11.6071068 & 41.9239372 & Background galaxy\\
17013 & 004627.00+420152.7 & 11.6125292 & 42.0313229 & Globular Cluster\\
17014 & 004627.00+420152.7 & 11.6125292 & 42.0313229 & Globular Cluster\\
17013 & 004630.89+420115.2 & 11.6286647 & 42.0208819 & Background galaxy\\
17014 & 004630.99+420956.1 & 11.6291078 & 42.1655802 & Background galaxy
\enddata
\label{aligntab}
\end{deluxetable}

\clearpage


\clearpage

\begin{deluxetable}{ll}
\tablewidth{7.0in}
\tablecaption{Chandra PHAT Catalog column names and descriptions}.
\tablehead{
\colhead{Column} &
\colhead{Description}
}
\startdata
Src & Source number in this catalog\\
ObsID & Observation in which the source was measured\\
Catalog Name (CXO~J) & Unique source identifier from coordinates. Prefix for all is CXO~J. \\
RA & J2000 Right Ascension\\
DEC & J2000 Declination\\
$\sigma$ [$''$] & X-ray position uncertainty in arcsec\\
$\Theta$ [$'$] & Off-axis angle in the Chandra observation in arcmin\\
Cts 0.35-8.0 & Source Net Counts in the 0.35-8.0 keV band\\
Cts 0.35-1.0 & Source Net Counts in the 0.35-2.0 keV band\\
Cts 0.5-8.0 &  Source Net Counts in the 0.5-8.0 keV band\\
Cts 0.5-1.0 &  Source Net Counts in the 0.5-1.0 keV band\\
Cts 1.0-2.0 &  Source Net Counts in the 1.0-2.0 keV band\\
Cts 2.0-4.0 &  Source Net Counts in the 2.0-4.0 keV band\\
Cts 2.0-8.0 &  Source Net Counts in the 2.0-8.0 keV band\\
Cts 4.0-8.0 &  Source Net Counts in the 4.0-8.0 keV band\\
Rate 0.35-8.0 & ARF-corrected Count Rate in the 0.35-8.0 keV band\\
Rate 0.35-1.0 & ARF-corrected Count Rate in the 0.35-2.0 keV band\\
Rate 0.5-8.0 & ARF-corrected Count Rate in the 0.5-8.0 keV band\\
Rate 0.5-1.0 &  ARF-corrected Count Rate in the 0.5-1.0 keV band\\
Rate 1.0-2.0 &  ARF-corrected Count Rate in the 1.0-2.0 keV band\\
Rate 2.0-4.0 &  ARF-corrected Count Rate in the 2.0-4.0 keV band\\
Rate 2.0-8.0 &  ARF-corrected Count Rate in the 2.0-8.0 keV band\\
Rate 4.0-8.0 & ARF-corrected Count Rate in the 4.0-8.0 keV band\\
PNS 0.35-8.0 & The probability that no source was at this location in the 0.35-8.0 keV band\\
Flux 0.35-8.0 & Energy flux (erg cm$^{-2}$ s$^{-1}$) in the 0.35-8.0 keV band (Count Rate * 1.313E-11)\\
Flux 0.35-1.0 & Energy flux (erg cm$^{-2}$ s$^{-1}$) in the 0.35-1.0 keV band (Count Rate * 1.767E-11)\\
Flux 0.5-8.0 & Energy flux (erg cm$^{-2}$ s$^{-1}$) in the 0.5-8.0 keV band (Count Rate * 1.272E-11)\\
Flux 0.5-1.0 & Energy flux (erg cm$^{-2}$ s$^{-1}$) in the 0.5-1.0 keV band (Count Rate * 1.438E-10)\\
Flux 1.0-2.0 & Energy flux (erg cm$^{-2}$ s$^{-1}$) in the 1.0-2.0 keV band (Count Rate * 5.620E-12)\\
Flux 2.0-4.0 & Energy flux (erg cm$^{-2}$ s$^{-1}$) in the 2.0-4.0 keV band (Count Rate * 1.443E-11)\\
Flux 2.0-8.0 & Energy flux (erg cm$^{-2}$ s$^{-1}$) in the 2.0-8.0 keV band (Count Rate * 2.138E-11)\\
Flux 4.0-8.0 & Energy flux (erg cm$^{-2}$ s$^{-1}$) in the 4.0-8.0 keV band (Count Rate * 3.487E-11)\\
Flux 0.2-4.5 & Energy Flux (erg cm$^{-2}$ s$^{-1}$) in the 0.2-4.5 keV band (0.35-8.0 keV Count Rate * 9.416E-12) \\  
XMM ID & Matched source identifier in S11\\
XMM Class & Source classification in S11\\
PHAT & Single letter PHAT counterpart code: \\
  & n (none), g (galaxy), p (point/star), f (foreground), s (SNR), c (cluster)\\
$A_{\rm V}$ & Mean extinction from \citet{dalcanton2015} maps\\
Notes & Notes from visual inspection of the PHAT images\\
\enddata
\label{columns}
\end{deluxetable}

\clearpage

\begin{center}
\begin{table*}
\caption{Chandra PHAT X-ray Catalog. 0.2-4.5 keV fluxes were calculated from net counts and exposure and converted to XMM-Newton comparable bands with galactic $N_{H}$ and $\Gamma$ values from S11 using PIMMS.  Full version available in machine-readable format only.}
\begin{tabular}{|c c c c c c c c c c c c}
\cline{1-12}
\cline{1-12}
Src & ObsID & Catalog Name & RA & DEC & $\sigma$ [$''$] & $\Theta$ [$'$] & Cts 0.35-8.0 & Cts 0.35-1.0 & Cts 0.5-8.0 & Cts 0.5-1.0 &\\
\cline{1-12}
1 & 17009 & 004320.03+413645.8 & 10.833453 & 41.612731 & 3.7 & 8.51 & 12.0$_{-4.2}^{+5.4}$ & 2.3$_{-1.6}^{+2.9}$ & 12.0$_{-4.2}^{+5.4}$ & 2.5$_{-1.6}^{+2.9}$ & \\
2 & 17009 & 004323.63+413144.8 & 10.848449 & 41.529103 & 1.1 & 8.12 & 73.0$_{-8.8}^{+9.9}$ & 1.7$_{-1.3}^{+2.7}$ & 73.0$_{-8.8}^{+9.9}$ & 1.8$_{-1.3}^{+2.7}$ &\\
3 & 17009 & 004325.01+413554.6 & 10.854213 & 41.598498 & 2.2 & 7.42 & 15.0$_{-4.3}^{+5.5}$ & 0.56$_{-0.84}^{+2.3}$ & 15.0$_{-4.3}^{+5.5}$ & 0.73$_{-0.83}^{+2.3}$ &\\
\cline{1-12}
\end{tabular}
\label{catalog}
\end{table*}

\begin{table*}
\begin{tabular}{c c c c c c c c c}
\cline{1-9}
\cline{1-9}
 Cts 1.0-2.0 & Cts 2.0-4.0 & Cts 2.0-8.0 & Cts 4.0-8.0 & Rate 0.35-8.0 & Rate 0.35-1.0 & Rate 0.5-8.0 & Rate 0.5-1.0 & \\
\cline{1-9}
 6.9$_{-2.8}^{+4.0}$ & 1.6$_{-1.6}^{+3.0}$ & 2.8$_{-2.6}^{+3.8}$ & 1.2$_{-1.9}^{+3.2}$ & 2.4e-04$_{-8.6e-05}^{+1.1e-04}$ & 2.3e-04$_{-1.6e-04}^{+3.0e-04}$ & 2.3e-04$_{-8.2e-05}^{+1.0e-04}$ & 2.1e-04$_{-1.4e-04}^{+2.5e-04}$ &\\
  28.0$_{-5.4}^{+6.5}$ & 29.0$_{-5.5}^{+6.6}$ & 43.0$_{-6.8}^{+7.9}$ & 14.0$_{-4.0}^{+5.1}$ & 1.5e-03$_{-1.8e-04}^{+2.1e-04}$ & 2.5e-04$_{-1.9e-04}^{+4.0e-04}$ & 1.5e-03$_{-1.8e-04}^{+2.0e-04}$ & 2.3e-04$_{-1.6e-04}^{+3.4e-04}$ &\\
  6.1$_{-2.6}^{+3.8}$ & 2.0$_{-1.6}^{+2.9}$ & 8.5$_{-3.3}^{+4.4}$ &  6.5$_{-2.8}^{+4.0}$ & 3.2e-04$_{-9.2e-05}^{+1.2e-04}$ & 5.3e-05$_{-8.0e-05}^{+2.2e-04}$ & 3.1e-04$_{-8.8e-05}^{+1.1e-04}$ & 5.9e-05$_{-6.8e-05}^{+1.9e-04}$ &\\
\cline{1-9}

\end{tabular}
\end{table*}

\begin{table*}
\begin{tabular}{c c c c c c c c}
\cline{1-8}
\cline{1-8}
 Rate 1.0-2.0 & Rate 2.0-4.0 & Rate 2.0-8.0 & Rate 4.0-8.0 & PNS 0.35-8.0 & Flux 0.35-8.0 &  Flux 0.35-1.0 & \\
\cline{1-8}
   1.5e-04$_{-5.9e-05}^{+8.4e-05}$ & 4.1e-05$_{-4.1e-05}^{+7.3e-05}$ & 9.6e-05$_{-8.9e-05}^{+1.3e-04}$ & 3.1e-05$_{-5.0e-05}^{+8.3e-05}$ & 1.1e-04 &  3.2e-15$^{+1.4e-15}_{-1.1e-15}$ & 4e-15$^{+5.2e-15}_{-2.9e-15}$ &\\
 6.6e-04$_{-1.3e-04}^{+1.5e-04}$ & 7.4e-04$_{-1.4e-04}^{+1.7e-04}$ & 1.5e-03$_{-2.3e-04}^{+2.7e-04}$ & 3.5e-04$_{-1.0e-04}^{+1.3e-04}$ & 0.0e+00 &  2e-14$^{+2.7e-15}_{-2.4e-15}$ & 4.5e-15$^{+7e-15}_{-3.4e-15}$ &\\
 1.3e-04$_{-5.7e-05}^{+8.4e-05}$ & 5.3e-05$_{-4.3e-05}^{+7.8e-05}$ & 3.1e-04$_{-1.2e-04}^{+1.6e-04}$ & 1.8e-04$_{-7.5e-05}^{+1.1e-04}$ & 6.3e-08 & 4.2e-15$^{+1.5e-15}_{-1.2e-15}$ &  9.4e-16$^{+3.9e-15}_{-1.4e-15}$ &\\
\cline{1-8}
\end{tabular}
\end{table*}

\begin{table*}
\begin{tabular}{c c c c c c c}
\cline{1-7}
\cline{1-7}
  Flux 0.5-8.0 & Flux 0.5-1.0 & Flux 1.0-2.0 & Flux 2.0-4.0 & Flux 2.0-8.0 & Flux 4.0-8.0 & Flux 0.2-4.5\\
\cline{1-7}
 3e-15$^{+1.3e-15}_{-1e-15}$ & 3e-15$^{+3.6e-15}_{-2e-15}$ & 8.2e-16$^{+4.7e-16}_{-3.3e-16}$ & 5.9e-16$^{+1.1e-15}_{-5.9e-16}$ & 2e-15$^{+2.8e-15}_{-1.9e-15}$ & 1.1e-15$^{+2.9e-15}_{-1.8e-15}$ & $2.8e-15^{+1e-15}_{-1e-15}$ \\
1.9e-14$^{+2.5e-15}_{-2.2e-15}$ & 3.3e-15$^{+4.9e-15}_{-2.4e-15}$ & 3.7e-15$^{+8.6e-16}_{-7.1e-16}$ & 1.1e-14$^{+2.4e-15}_{-2e-15}$ & 3.2e-14$^{+5.8e-15}_{-5e-15}$ & 1.2e-14$^{+4.6e-15}_{-3.6e-15}$ & $1.7e-14^{+2e-15}_{-2e-15}$ \\
4e-15$^{+1.4e-15}_{-1.1e-15}$ &  8.5e-16$^{+2.7e-15}_{-9.7e-16}$ & 7.6e-16$^{+4.7e-16}_{-3.2e-16}$ & 7.7e-16$^{+1.1e-15}_{-6.2e-16}$ & 6.5e-15$^{+3.4e-15}_{-2.5e-15}$ & 6.1e-15$^{+3.7e-15}_{-2.6e-15}$ & $3.7e-15^{+1e-15}_{-1e-15}$ \\
\cline{1-7}
\end{tabular}
\end{table*}

\begin{table*}
\begin{tabular}{c c c c c|}
\cline{1-5}
\cline{1-5}
S11 & S11 Class & PHAT & $A_{\rm V}$ & Notes\\
\cline{1-5}
1197 & $\langle$hard$\rangle$ & n & \nodata & \\
1213 & $\langle$hard$\rangle$ & g & 0.63 &  AP galaxy (XMM-Newton HR agrees) \\
1217 & $\langle$hard$\rangle$ & n & \nodata & \\
\cline{1-5}

\end{tabular}
\end{table*}
\end{center}

\begin{table*}
\caption{Supernova Remnants Detected in this Survey}
\label{tab:snrs}
\begin{center}
\begin{tabular}{ l c c c c c c c c}
\cline{1-9} 
\cline{1-9} 
Catalog Name & SPH11~\# & Optical/Radio & L\&L\# & Counts & L(0.35-8.0 keV)               
& Minimum off-axis  & Extended & Hard \\ 
            &          &               & & (0.35-8.0 keV) & $(\mathrm{10^{35}~erg~s^{-1}})$       
& angle (arc min)   &    &       \\
\hline 
004624.58+415543.5   & 1793   & Radio \citep{braun1990}	&     & 13.6$\pm4.8$  &	$2.6\pm1.1$  & 3.51		  & Unclear & No \\
004452.84+415459.7   & 1539   & Optical & 112 & 19.2$\pm5.0$   & $3.2\pm1.0$             & 3.51		  & Unclear & No  \\
004513.88+413615.7   & 1599   & Optical & 124 & 83.1$\pm11.1$ & $13.4\pm1.7$             & 5.03		  & Yes     & No \\
004339.24+412653.2   & 1275   & Optical &  70 & 24.8$\pm5.7$  & $9.7\pm2.0$  & 7.74		  & Unclear & No \\
004413.49+411954.1  & 1410   & Optical &  87 & 40.5$\pm7.2$  &	$9.7\pm1.4$               & 3.39		  & Yes     & No \\
004451.06+412906.6   & 1535   & Optical & 111 & 76.8$\pm12.2$  &	$8.2\pm1.6$\tablenotemark{a}               & 3.31		  & Yes     & Yes \\
\cline{1-9}
\end{tabular}
\begin{tablenotes}
    \item[a]{$^{\rm a}$} Spectral model is {\tt APEC} plus a power-law.
\end{tablenotes}
\end{center}
\end{table*}

\clearpage

\begin{threeparttable}
\caption{PHAT Point Source Counterpart Candidate Multiwavelength Properties}
\begin{center}
\tiny
\begin{tabular}{ccccccccccccc}
\cline{1-13}
Name (CXO~J) & CXO RA & CXO Dec & $\sigma$\tablenotemark{b} ($''$) & PHAT RA & PHAT Dec & F275W & F336W & F475W & F814W & F110W & F160W & Flux 0.35-8.0\\
\cline{1-13}
004339.06+412117.6\tablenotemark{a} & 10.912737 & 41.354885 & 0.66 & 10.912961 & 41.354867 & 23.84 & 23.23 & 23.87 & 23.63 & 23.60 & 23.27 & $4.9e-14^{+4e-15}_{-4e-15}$\\
004420.18+413408.2\tablenotemark{a} & 11.084068 & 41.568957 & 0.55 & 11.084035 & 41.568901 & 21.60 & 21.59 & 22.58 & 22.19 & 21.89 & 21.67 & $4.4e-15^{+1e-15}_{-1e-15}$\\
004445.88+413152.3\tablenotemark{a} & 11.191148 & 41.531182 & 4.6 & 11.191047 & 41.531859 & 21.58 & 21.28 & 22.30 & 21.55 & 21.17 & 21.00 & $6.2e-15^{+2e-15}_{-2e-15}$\\
004514.76+415034.5\tablenotemark{a} & 11.311509 & 41.842926 & 0.77 & 11.311492 & 41.842875 & 18.94 & 19.17 & 20.57 & 20.49 & 99.999 & 99.999 & $1e-14^{+2e-15}_{-2e-15}$\\
004536.13+414702.5\tablenotemark{a} & 11.400529 & 41.784043 & 0.81 & 11.400650 & 41.784065 & 19.31 & 19.27 & 20.38 & 19.81 & 19.76 & 19.64 & $3.2e-15^{+1e-15}_{-1e-15}$\\
004537.67+415124.4\tablenotemark{a} & 11.406964 & 41.856792 & 2.3 & 11.407024 & 41.856355 & 20.60 & 20.46 & 21.64 & 21.08 & 20.95 & 20.82 & $5.9e-15^{+2e-15}_{-2e-15}$\\
004637.22+421034.5\tablenotemark{a} & 11.655092 & 42.176248 & 0.75 & 11.655076 & 42.176184 & 18.63 & 19.00 & 20.54 & 20.53 & 20.70 & 20.70 & $3.1e-15^{+1e-15}_{-1e-15}$\\
004639.47+420649.2\tablenotemark{a} & 11.664470 & 42.113665 & 0.89 & 11.664686 & 42.113558 & 23.70 & 23.69 & 24.33 & 24.24 & 25.71 & 25.50 & $3.7e-15^{+1e-15}_{-1e-15}$\\
\hline
004350.76+412118.1\tablenotemark{c} & 10.961516 & 41.355033 & 0.45 & 10.961508 & 41.355045 & 22.78 & 21.16 & 21.69 & 19.84 & 19.13 & 18.25 & $4.4e-14^{+4e-15}_{-4e-15}$\\
004352.37+412222.8\tablenotemark{c} & 10.968197 & 41.372997 & 0.9 & 10.968015 & 41.373104 & 24.50 & 23.10 & 23.30 & 21.90 & 21.43 & 20.83 & $4.2e-15^{+2e-15}_{-1e-15}$\\
004356.78+413410.9\tablenotemark{c} & 10.986584 & 41.569705 & 0.4 & 10.986615 & 41.569631 & 99.999 & 24.75 & 24.65 & 23.15 & 22.33 & 21.32 & $4.1e-15^{+1e-15}_{-1e-15}$\\
004402.02+414028.8\tablenotemark{c} & 11.008427 & 41.674679 & 1.1 & 11.008576 & 41.674590 & 22.10 & 21.10 & 22.13 & 20.69 & 20.05 & 19.36 & $7.3e-15^{+2e-15}_{-2e-15}$\\
004404.55+413159.4\tablenotemark{c} & 11.018958 & 41.533179 & 0.52 & 11.018864 & 41.533169 & 25.89 & 24.14 & 24.03 & 23.17 & 22.34 & 21.74 & $3.9e-15^{+1e-15}_{-1e-15}$\\
004407.44+412460.0\tablenotemark{c} & 11.031002 & 41.416662 & 0.64 & 11.031037 & 41.416678 & 24.86 & 23.06 & 20.93 & 17.31 & 16.20 & 15.49 & $1.8e-15^{+1e-15}_{-7e-16}$\\
004437.96+414512.6\tablenotemark{c} & 11.158149 & 41.753505 & 0.85 & 11.158152 & 41.753489 & 22.95 & 22.84 & 22.95 & 22.32 & 21.45 & 20.45 & $1.8e-14^{+3e-15}_{-3e-15}$\\
004452.51+411710.7\tablenotemark{c} & 11.218808 & 41.286303 & 1.2 & 11.218816 & 41.286414 & 23.15 & 19.75 & 20.88 & 19.83 & 19.42 & 18.48 & $3.5e-14^{+4e-15}_{-4e-15}$\\
004454.75+411918.3\tablenotemark{c} & 11.228112 & 41.321761 & 1.5 & 11.228090 & 41.321657 & 21.11 & 20.92 & 22.41 & 21.29 & 20.62 & 19.50 & $1.2e-14^{+3e-15}_{-2e-15}$\\
004525.67+413158.2\tablenotemark{c} & 11.356948 & 41.532845 & 1.1 & 11.356959 & 41.532850 & 23.26 & 22.47 & 22.56 & 21.04 & 20.25 & 19.47 & $3.6e-15^{+1e-15}_{-1e-15}$\\
004542.25+420817.9\tablenotemark{c} & 11.426039 & 42.138318 & 1.2 & 11.426153 & 42.138378 & 25.84 & 23.30 & 23.60 & 22.38 & 22.01 & 21.38 & $1.1e-14^{+2e-15}_{-2e-15}$\\
004552.94+420234.0\tablenotemark{c} & 11.470589 & 42.042780 & 0.86 & 11.470609 & 42.042771 & 26.52 & 23.84 & 23.67 & 22.14 & 21.32 & 20.48 & $1.1e-14^{+2e-15}_{-2e-15}$\\
004558.04+420302.9\tablenotemark{c} & 11.491829 & 42.050818 & 0.7 & 11.491873 & 42.050830 & 21.00 & 20.97 & 21.89 & 21.23 & 20.59 & 19.76 & $3.1e-14^{+5e-15}_{-4e-15}$\\
004612.67+421027.8\tablenotemark{c} & 11.552784 & 42.174382 & 0.39 & 11.552790 & 42.174390 & 24.12 & 22.55 & 23.08 & 21.31 & 20.76 & 20.09 & $7.5e-15^{+2e-15}_{-2e-15}$\\
004640.59+415422.8\tablenotemark{c} & 11.669118 & 41.906343 & 0.6 & 11.669103 & 41.906427 & 23.59 & 22.18 & 22.44 & 20.25 & 19.43 & 18.49 & $5.5e-14^{+5e-15}_{-4e-15}$\\
004652.18+421505.8\tablenotemark{c} & 11.717403 & 42.251599 & 2.1 & 11.716865 & 42.251591 & 22.42 & 22.03 & 23.15 & 21.75 & 21.38 & 20.19 & $6.8e-15^{+2e-15}_{-2e-15}$\\
004703.82+420453.0\tablenotemark{c} & 11.765906 & 42.081385 & 0.84 & 11.765933 & 42.081374 & 21.56 & 21.14 & 22.30 & 20.59 & 19.90 & 18.65 & $7.6e-14^{+6e-15}_{-5e-15}$\\
\hline
004336.08+413320.4 & 10.900328 & 41.555683 & 0.87 & 10.900361 & 41.555726 & 23.67 & 22.16 & 21.87 & 20.37 & 19.50 & 18.82 & $7.3e-15^{+2e-15}_{-1e-15}$\\
004357.54+413055.8 & 10.989765 & 41.515488 & 0.49 & 10.989700 & 41.515450 & 24.79 & 22.80 & 23.88 & 22.36 & 22.20 & 21.05 & $1.2e-14^{+2e-15}_{-2e-15}$\\
004359.83+412435.6 & 10.999304 & 41.409889 & 0.68 & 10.999142 & 41.410006 & 25.81 & 23.47 & 23.59 & 21.95 & 21.50 & 20.89 & $4.9e-15^{+2e-15}_{-1e-15}$\\
004412.04+413217.4 & 11.050183 & 41.538170 & 0.8 & 11.050257 & 41.538157 & 26.05 & 23.94 & 21.77 & 18.24 & 17.00 & 16.26 & $4.1e-15^{+3e-15}_{-2e-15}$\\
004412.17+413148.4 & 11.050698 & 41.530100 & 0.36 & 11.050679 & 41.530036 & 22.34 & 21.30 & 21.77 & 20.28 & 19.97 & 19.42 & $2.7e-14^{+3e-15}_{-3e-15}$\\
004413.18+412911.4 & 11.054915 & 41.486508 & 0.89 & 11.054808 & 41.486368 & 26.49 & 24.06 & 99.999 & 21.33 & 20.66 & 19.91 & $9.1e-15^{+2e-15}_{-2e-15}$\\
004422.57+414506.5 & 11.094057 & 41.751798 & 0.68 & 11.094097 & 41.751878 & 20.80 & 20.43 & 21.01 & 19.94 & 19.25 & 18.38 & $8e-14^{+6e-15}_{-5e-15}$\\
004424.80+413201.4 & 11.103325 & 41.533731 & 0.41 & 11.103325 & 41.533695 & 25.50 & 24.87 & 24.13 & 21.43 & 19.95 & 19.11 & $4.8e-14^{+4e-15}_{-4e-15}$\\
004425.73+412242.4 & 11.107221 & 41.378442 & 0.33 & 11.107175 & 41.378477 & 26.06 & 24.94 & 24.65 & 22.62 & 21.88 & 21.45 & $1.5e-14^{+2e-15}_{-2e-15}$\\
004431.82+415217.2 & 11.132597 & 41.871441 & 0.55 & 11.132547 & 41.871387 & 22.27 & 21.44 & 21.90 & 20.49 & 20.10 & 19.83 & $4.4e-15^{+2e-15}_{-1e-15}$\\
004448.13+412247.9 & 11.200545 & 41.379973 & 0.71 & 11.200584 & 41.380057 & 23.48 & 23.39 & 24.59 & 22.95 & 21.85 & 20.45 & $1.8e-14^{+3e-15}_{-2e-15}$\\
004453.33+415159.5 & 11.222218 & 41.866543 & 0.34 & 11.222263 & 41.866515 & 25.05 & 23.09 & 23.19 & 21.22 & 20.47 & 19.85 & $7.8e-15^{+2e-15}_{-1e-15}$\\
004455.72+415334.6 & 11.232187 & 41.892939 & 0.53 & 11.232176 & 41.892884 & 24.75 & 23.22 & 23.26 & 22.04 & 21.20 & 20.50 & $4.1e-15^{+2e-15}_{-1e-15}$\\
004459.11+414005.1 & 11.246280 & 41.668081 & 0.9 & 11.246206 & 41.668054 & 26.14 & 22.74 & 22.06 & 20.70 & 19.94 & 19.32 & $1.6e-14^{+3e-15}_{-2e-15}$\\
004500.89+414309.8 & 11.253707 & 41.719385 & 0.78 & 11.253848 & 41.719425 & 27.47 & 25.15 & 25.38 & 24.51 & 24.42 & 24.11 & $9.7e-15^{+2e-15}_{-2e-15}$\\
004502.33+414943.1 & 11.259688 & 41.828654 & 0.61 & 11.259761 & 41.828766 & 24.07 & 23.46 & 23.44 & 22.74 & 22.66 & 22.48 & $4.6e-15^{+1e-15}_{-1e-15}$\\
004510.96+414559.2 & 11.295681 & 41.766440 & 0.43 & 11.295651 & 41.766412 & 23.71 & 22.78 & 23.30 & 21.08 & 20.35 & 19.03 & $3.2e-14^{+3e-15}_{-3e-15}$\\
004526.67+415631.0 & 11.361109 & 41.941936 & 1.4 & 11.361074 & 41.942056 & 23.01 & 22.34 & 22.74 & 21.05 & 20.62 & 19.46 & $1.5e-14^{+3e-15}_{-3e-15}$\\
004527.88+413905.5 & 11.366179 & 41.651539 & 0.43 & 11.366180 & 41.651542 & 24.34 & 22.81 & 23.41 & 20.87 & 19.97 & 18.82 & $3e-14^{+3e-15}_{-3e-15}$\\
004528.24+412943.9 & 11.367681 & 41.495538 & 0.42 & 11.367687 & 41.495535 & 19.76 & 19.20 & 20.23 & 19.01 & 18.53 & 17.75 & $1.3e-13^{+7e-15}_{-6e-15}$\\
004537.84+414856.7 & 11.407660 & 41.815743 & 1.4 & 11.408020 & 41.815702 & 22.74 & 21.69 & 22.12 & 20.74 & 20.43 & 20.14 & $3.9e-15^{+2e-15}_{-1e-15}$\\
004543.15+415519.4 & 11.429790 & 41.922042 & 1.4 & 11.429859 & 41.922029 & 22.14 & 21.40 & 21.80 & 20.24 & 19.72 & 19.36 & $3.7e-15^{+2e-15}_{-1e-15}$\\
004550.83+415835.1 & 11.461787 & 41.976427 & 0.58 & 11.461834 & 41.976518 & 22.06 & 21.88 & 23.13 & 22.08 & 21.32 & 20.20 & $5.5e-15^{+2e-15}_{-1e-15}$\\
004607.50+420855.7 & 11.531257 & 42.148817 & 0.41 & 11.531258 & 42.148852 & 27.37 & 24.31 & 24.98 & 23.27 & 22.69 & 21.88 & $9.5e-15^{+2e-15}_{-2e-15}$\\
004611.38+415903.9 & 11.547414 & 41.984406 & 0.31 & 11.547478 & 41.984419 & 99.999 & 24.19 & 24.81 & 23.01 & 22.29 & 21.18 & $2.1e-14^{+3e-15}_{-3e-15}$\\
004611.85+420827.9 & 11.549374 & 42.141082 & 0.28 & 11.549385 & 42.141077 & 99.999 & 24.77 & 23.50 & 20.70 & 19.21 & 18.31 & $7.2e-14^{+5e-15}_{-5e-15}$\\
004613.49+415043.3 & 11.556209 & 41.845353 & 0.71 & 11.556171 & 41.845410 & 20.34 & 20.12 & 21.20 & 19.94 & 19.60 & 18.60 & $3.9e-14^{+4e-15}_{-4e-15}$\\
004617.57+415913.6 & 11.573194 & 41.987108 & 0.42 & 11.573203 & 41.987150 & 22.54 & 21.65 & 22.85 & 21.55 & 21.03 & 20.48 & $7.2e-15^{+2e-15}_{-1e-15}$\\
004630.46+421028.7 & 11.626900 & 42.174642 & 0.43 & 11.626948 & 42.174642 & 24.37 & 22.99 & 22.91 & 21.10 & 20.56 & 19.90 & $5e-15^{+2e-15}_{-1e-15}$\\
004630.68+420947.0 & 11.627815 & 42.163055 & 0.39 & 11.627821 & 42.163064 & 25.10 & 24.55 & 26.15 & 24.02 & 22.66 & 21.45 & $7.3e-15^{+2e-15}_{-2e-15}$\\
004648.19+420855.4 & 11.700785 & 42.148718 & 0.42 & 11.700789 & 42.148730 & 22.34 & 21.65 & 21.84 & 19.93 & 19.28 & 18.29 & $5.7e-14^{+4e-15}_{-4e-15}$\\
004648.27+420851.1 & 11.701118 & 42.147536 & 0.55 & 11.701051 & 42.147534 & 22.75 & 21.88 & 22.35 & 20.76 & 20.32 & 19.24 & $2.1e-14^{+3e-15}_{-3e-15}$\\
\cline{1-13}
\end{tabular}
\begin{tablenotes}
    \item[a] Included in the ``best candidate'' sample. CXO~J004339.06+412117.6 is considered such a candidate, but no SFH is available.
    \item[b] X-ray source position uncertainty in arcseconds from Table~\ref{catalog}.
    \item[c] No local young population detected.
\end{tablenotes}
\end{center}
\label{phat_photometry}
\end{threeparttable}

\clearpage

\begin{center}
\begin{threeparttable}
\caption{Best HMXB Candidate Age Probabilities: full table electronic only}
\begin{tabular}{cccccc}
\cline{1-6}
Catalog Name & Low Age (Myr)\tablenotemark{a} & High Age (Myr) & Prob.\tablenotemark{b} & $+$err $-$err\\
\cline{1-6}
004425.73+412242.4, & 4.0 & 5.0 & 0.000 & 0.077 & 0.000 \\
004425.73+412242.4, & 5.0 & 6.3 & 0.000 & 0.127 & 0.000 \\
004425.73+412242.4, & 6.3 & 7.9 & 0.000 & 0.182 & 0.000 \\
004425.73+412242.4, & 7.9 & 10.0 & 0.000 & 0.254 & 0.000 \\
004425.73+412242.4, & 10.0 & 12.6 & 0.795 & 0.000 & 0.760 \\
004425.73+412242.4, & 12.6 & 15.8 & 0.000 & 0.506 & 0.000 \\
004425.73+412242.4, & 15.8 & 20.0 & 0.000 & 0.460 & 0.000 \\
004425.73+412242.4, & 20.0 & 25.1 & 0.021 & 0.406 & 0.021 \\
004425.73+412242.4, & 25.1 & 31.6 & 0.000 & 0.403 & 0.000 \\
004425.73+412242.4, & 31.6 & 39.8 & 0.000 & 0.409 & 0.000 \\
004425.73+412242.4, & 39.8 & 50.1 & 0.000 & 0.414 & 0.000 \\
004425.73+412242.4, & 50.1 & 63.1 & 0.000 & 0.405 & 0.000 \\
004425.73+412242.4, & 63.1 & 79.4 & 0.184 & 0.136 & 0.184 \\
004448.13+412247.9, & 4.0 & 5.0 & 0.000 & 0.098 & 0.000 \\
004448.13+412247.9, & 5.0 & 6.3 & 0.000 & 0.160 & 0.000 \\
004448.13+412247.9, & 6.3 & 7.9 & 0.000 & 0.232 & 0.000 \\
004448.13+412247.9, & 7.9 & 10.0 & 0.000 & 0.304 & 0.000 \\
\nodata & \nodata & \nodata & \nodata & \nodata & \nodata\\
\cline{1-6}
\end{tabular}
\begin{tablenotes}
    \item[a] Low and high ages refer to the edges of the age bin.  For example, if low age is 4.0 and high age is 5.0, then the Prob. column refers to the probability that the HMXB candidate has an age between 4.0 and 5.0 Myr.
    \item[b] Probability that the HMXB system has an age in this interval.  There are uncertainties on this probability, given by the +err and -err columns.
\end{tablenotes}
\label{HMXB_ages}
\end{threeparttable}

\end{center}


\clearpage

\clearpage

\begin{figure*}
\begin{center}
\includegraphics[width=5.0in]{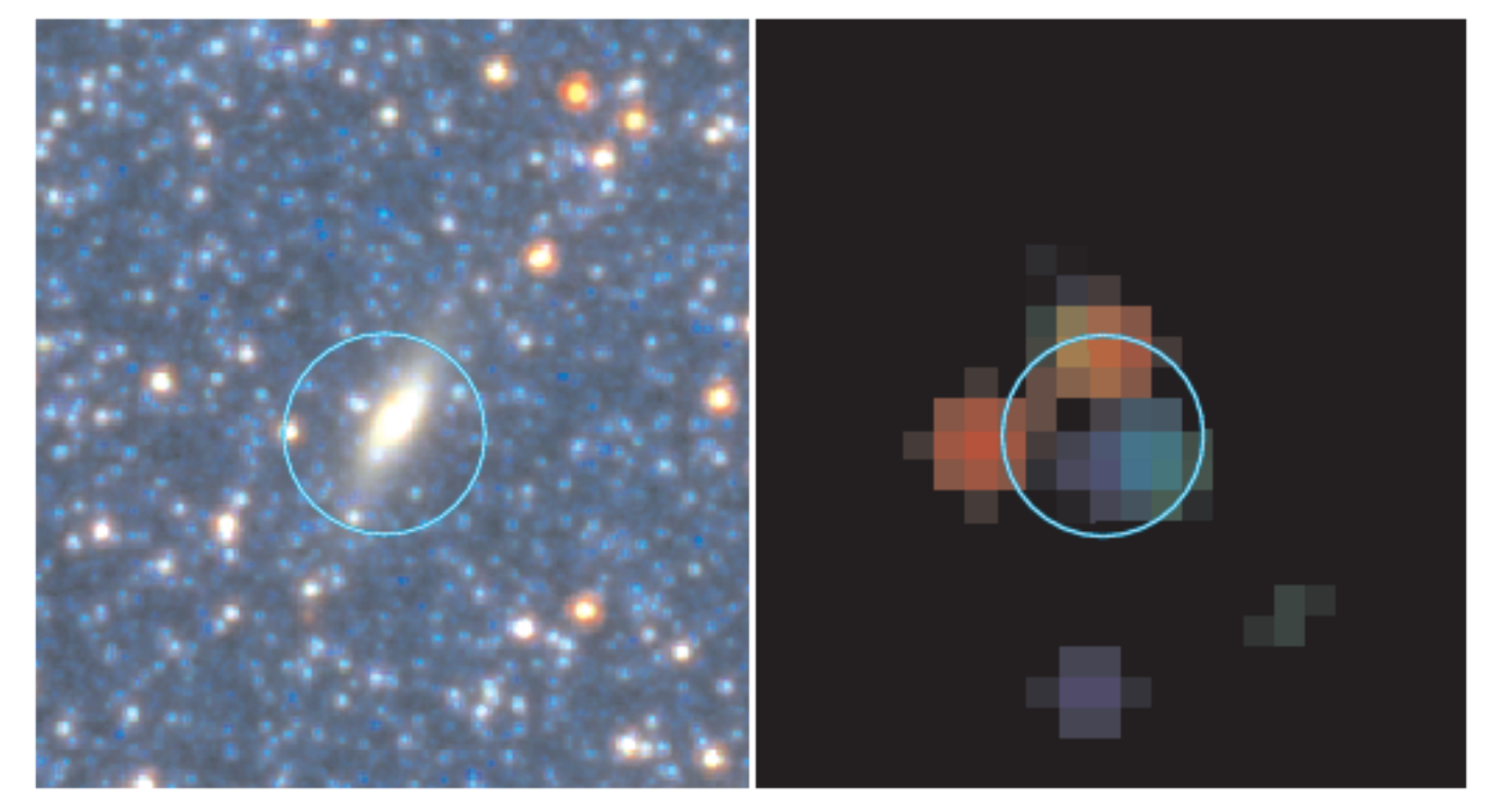}
\includegraphics[width=5.0in]{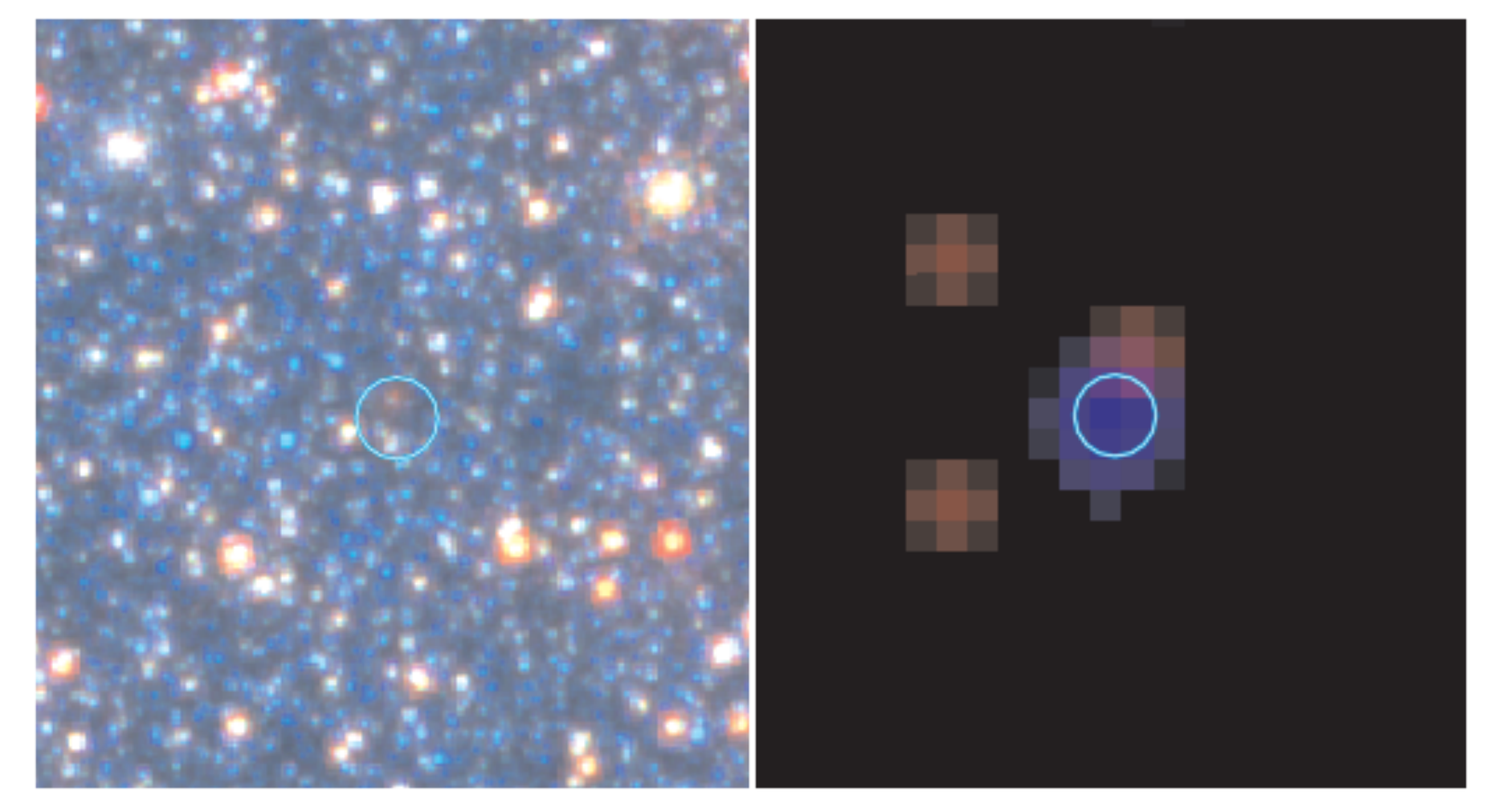}
\end{center}
\caption{Same structure as the panels of Figure~\ref{alignment}, but
  instead of alignment sources, these are examples of background
  galaxies found in the PHAT imaging after alignment.  {\it Top:}
  Source CXO~J004535.86+413322.8 in observation 17010 is clearly
  extended and red in the optical.  The left panel shows the PHAT
  image (red=F160W, green=F814W, blue=F475W), with the {\it Chandra}
  1-$\sigma$ position error marked in blue.  The right panel shows the
  same markings on the color {\it Chandra} image (red=0.35-1 keV,
  green=1-2 keV, blue=2-8 keV). The source is clearly extended and red
  in the optical.  {\it Bottom} Source CXO~J004437.52+415124.9 in
  observation 17012.  Panels show images and markings corresponding to
  the same bands and instruments as {\it top}. Here the background
  galaxy near the top of the 1-$\sigma$ position error is very faint
  and red, but distinct from the stars in the image.}
\label{AGN}
\end{figure*}

\begin{figure*}
\begin{center}
\includegraphics[width=6.0in]{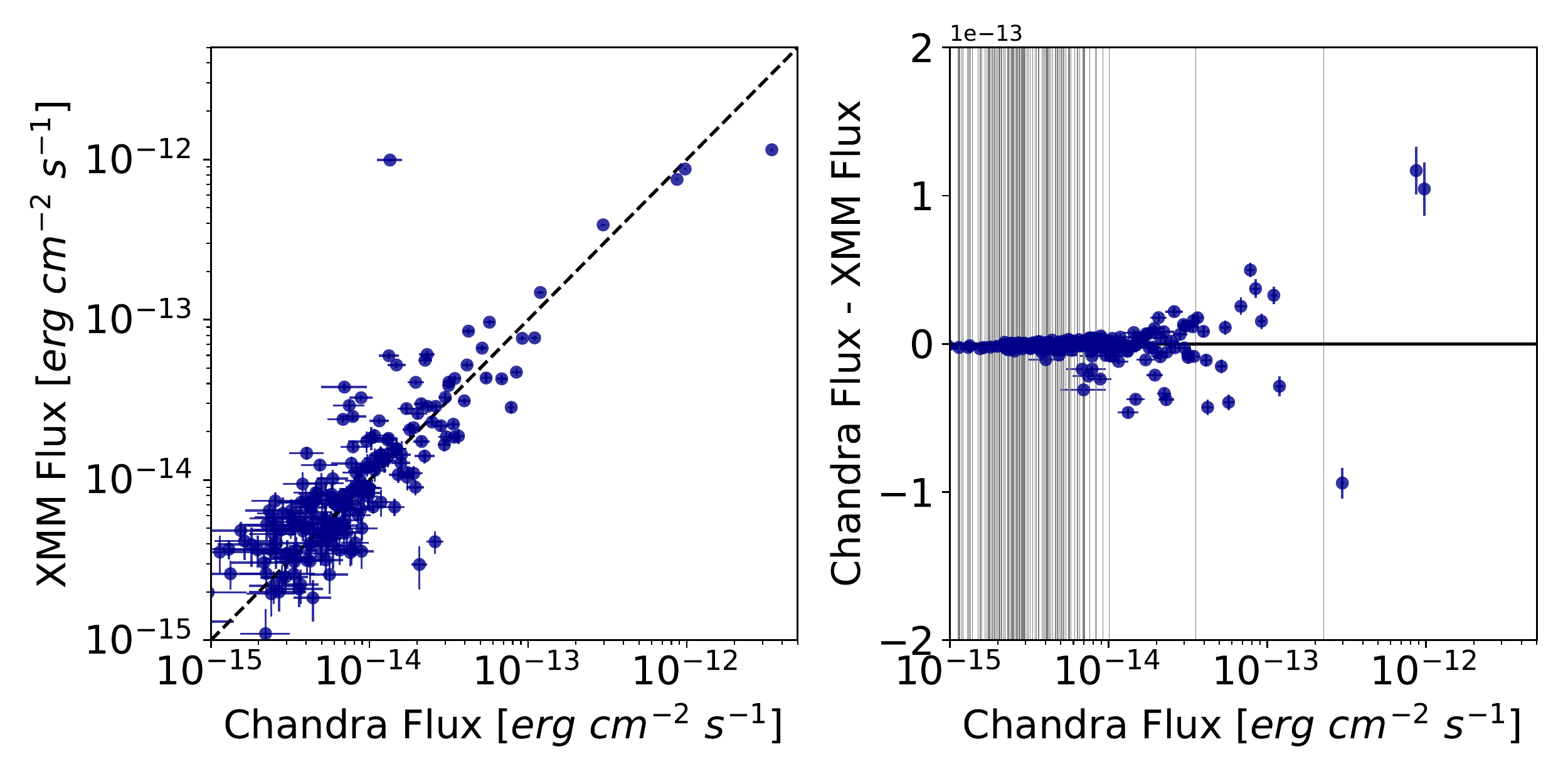}
\end{center}
\caption{Comparison between Chandra and XMM-Newton 0.5-4.0 keV fluxes, using the cross-matching from our catalog, assuming a power law spectrum with slope 1.7 and $N_H$=7$\times$10$^{20}$ cm$^{-2}$.  {\it Left:}  Direct comparison of fluxes between measurements for all sources matched. {\it Right:} Residuals of all matched sources, with sources undetected by XMM-Newton marked with gray vertical lines.\label{xmm_compare}}
\end{figure*}

\begin{figure*}
\begin{center}
\includegraphics[width=5.0in]{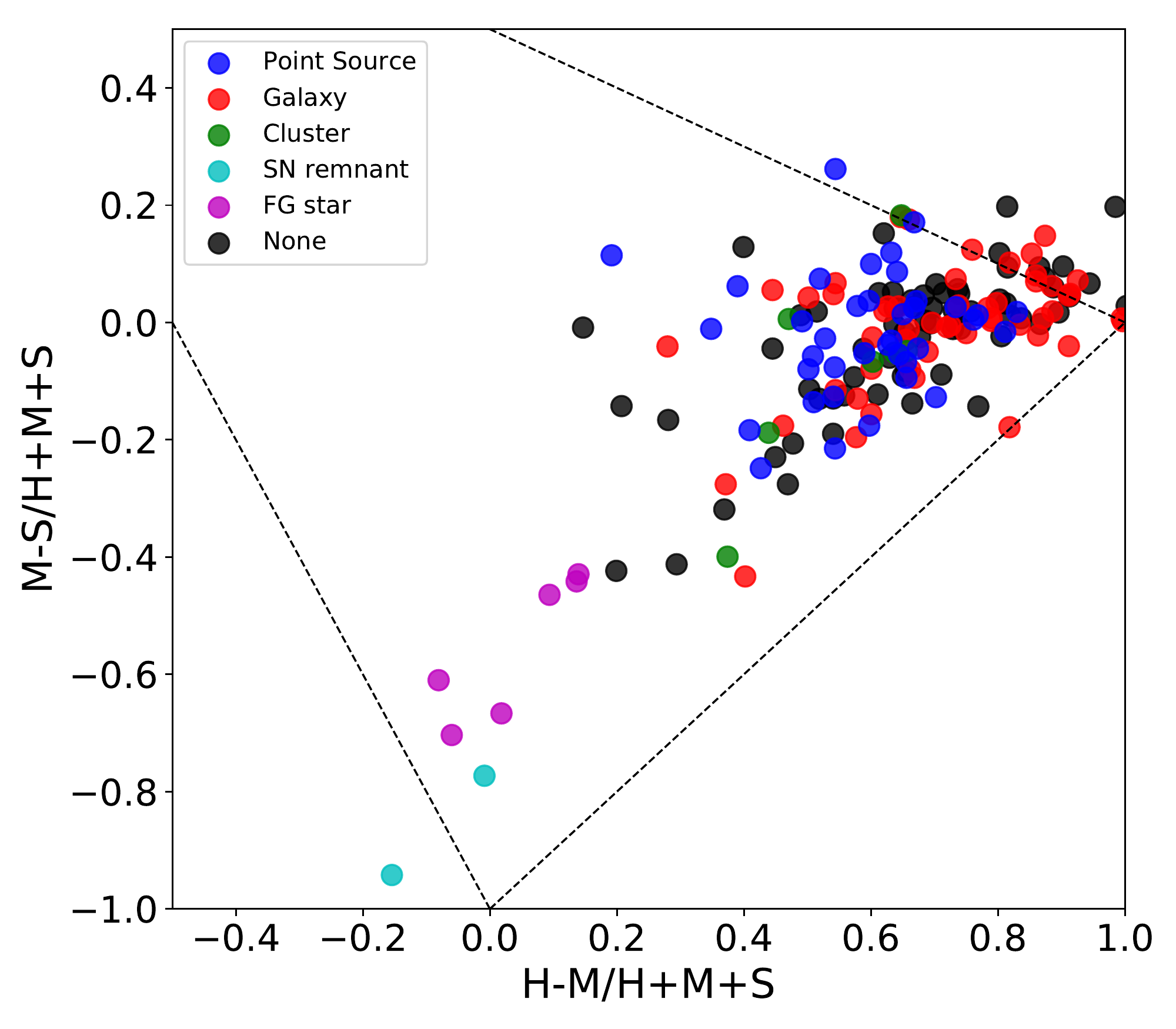}
\end{center}
\caption{Hardness ratios for all sources with $>$20 counts.  Bands are fluxes in: H=2-8~keV, M=1-2~keV, S=0.35-1~keV, and points are color-coded by their counterpart candidate type. The triangle marks the allowed region for positive counts in all bands.  Many of the hardest sources show negative counts in the soft band due to uncertainties in the background level when zero counts are detected in the soft band.  The area at the bottom of the plot is dominated by foreground stars and SNRs, and the area at the extreme right of the plot is dominated by background galaxies. \label{HRs}}
\end{figure*}

\clearpage

\begin{figure*}
\begin{center}
\includegraphics[width=3.0in]{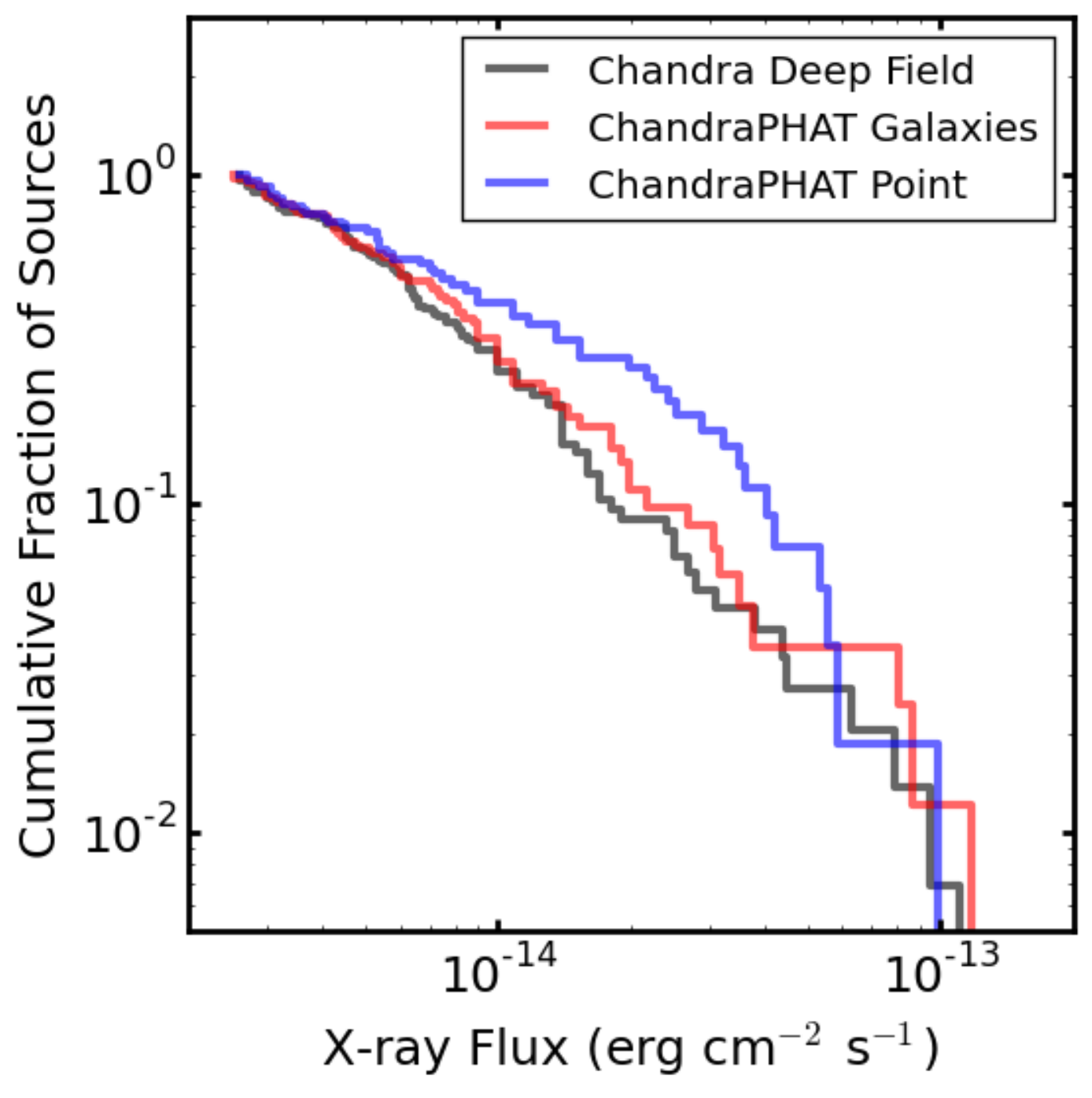}
\includegraphics[width=3.1in]{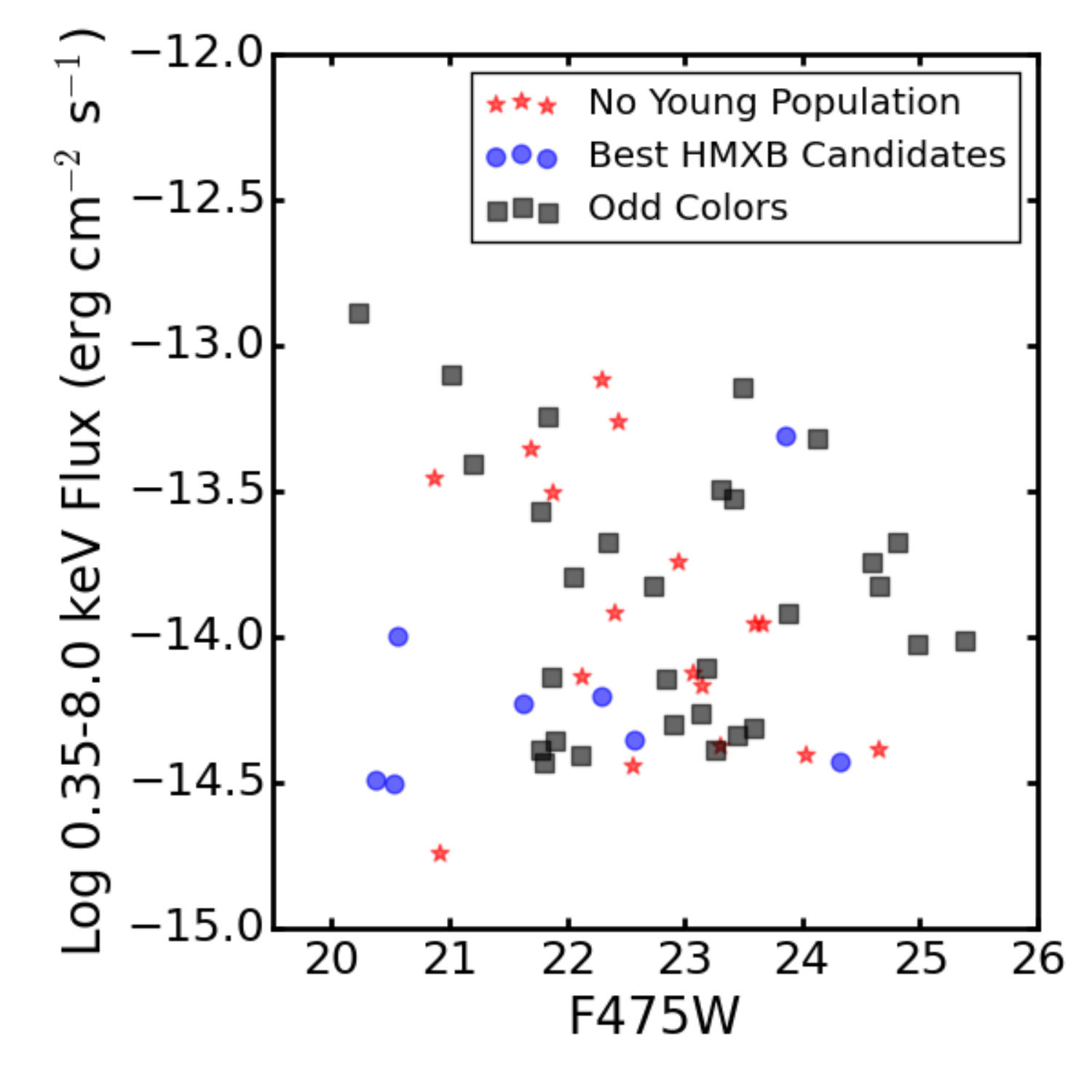}
\end{center}
\caption{{\it Left:} The fractional cumulative 0.5-7.0 keV flux histograms for the Chandra Deep Field (CDF) \citep{luo2017}, the 107 sources with background galaxy candidates in our sample, and the 58 sources with point source optical candidates in our sample.  While the galaxies follow a very similar distribution to the CDF, the point sources do not, confirming that they represent a different population of sources. {\it Right:} The X-ray fluxes vs. F475W magnitudes of the point source counterpart candidates.  The different point types correspond to the three tiers of counterpart candidates in Table~\ref{phat_photometry}.  \label{xlfs}}

\end{figure*}

\clearpage

\begin{figure*}
\begin{center}
\includegraphics[width=4.0in]{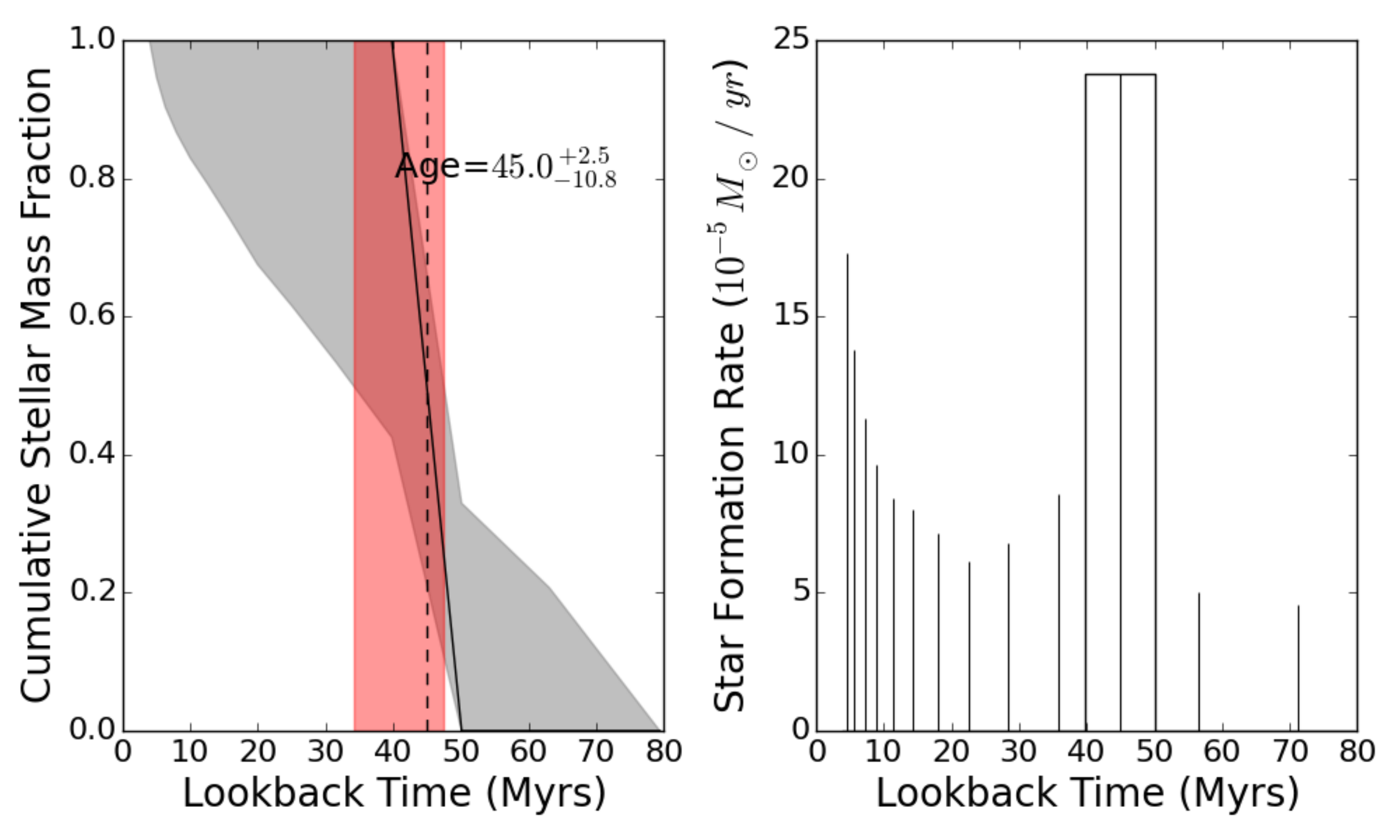}
\includegraphics[width=4.0in]{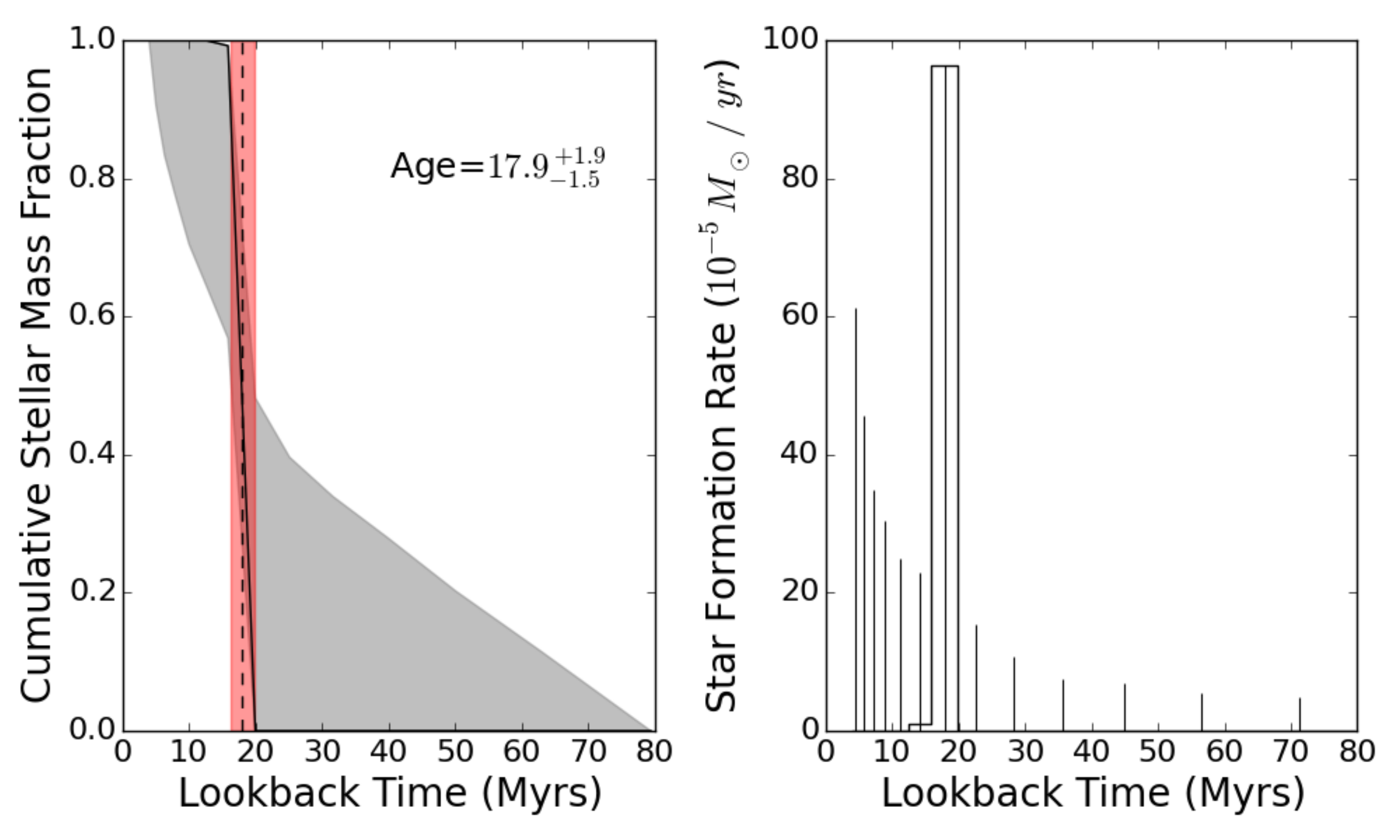}
\includegraphics[width=4.0in]{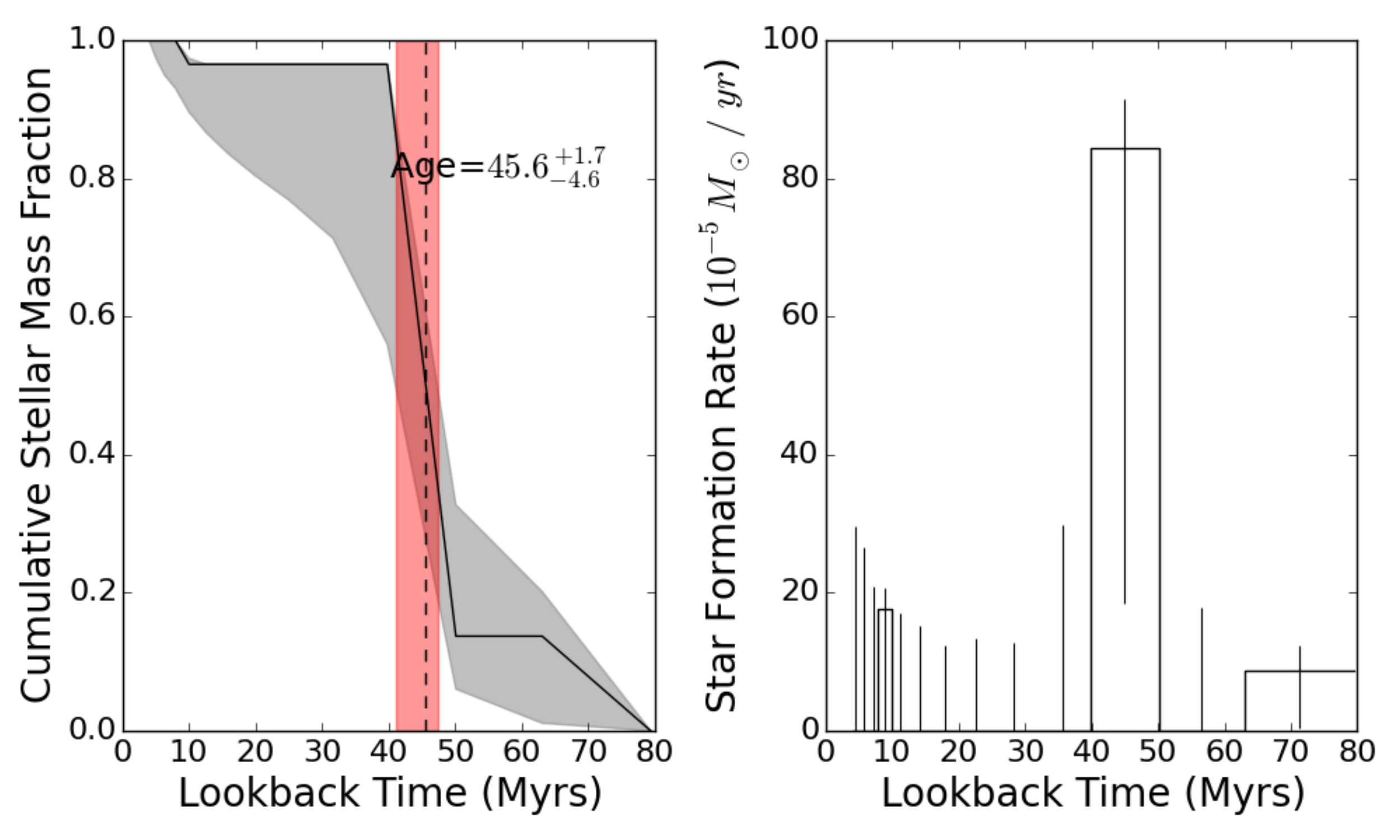}
\end{center}
\caption{Age estimation of HMXB candidates.  These ages refer to the time since the parent binary system formed, and do not probe how long the binary has been producing X-rays.   {\it Top:} CXO~J004420.20+413407.4. {\it Middle:} CXO~J004514.78+415034.2. {\it Bottom:} CXO~J004637.23+421033.7. {\it Left:} Cumulative stellar mass fraction as a function of age for all stars $<$80 Myr old.  This fractional distribution is calculated from the rates and uncertainties in {\it right}, which shows the most recent 80 Myr of the \citet{lewis2015} star formation history of the region where the HMXB is located. Dashed line shows the median age for the best fit distribution. The gray shading shows the 1$\sigma$ uncertainty range, and the red shading shows the ages consistent with the median age within the uncertainties.}
\label{age_examples}
\end{figure*}


\begin{figure*}
\includegraphics[width=2.8in]{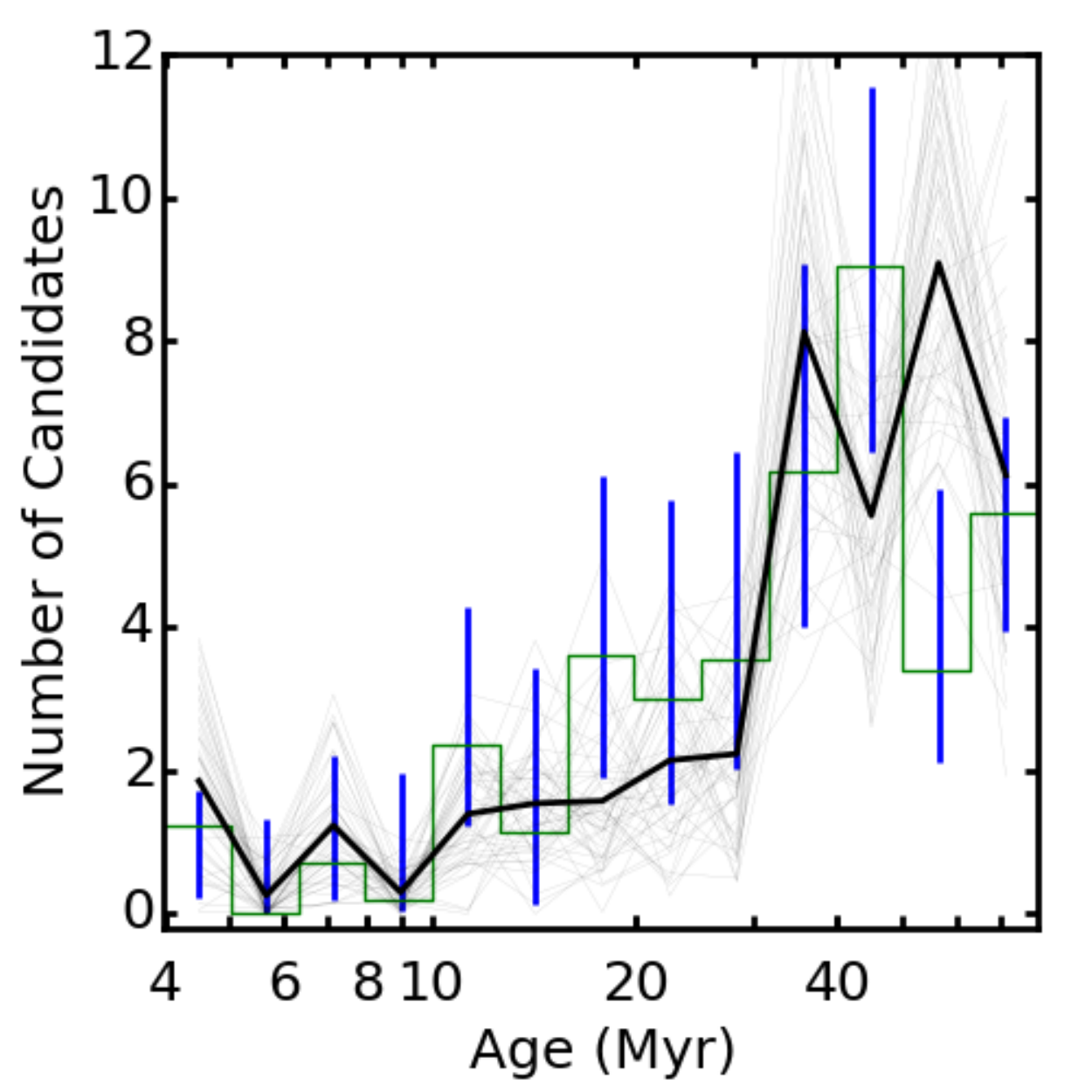}
\includegraphics[width=2.8in]{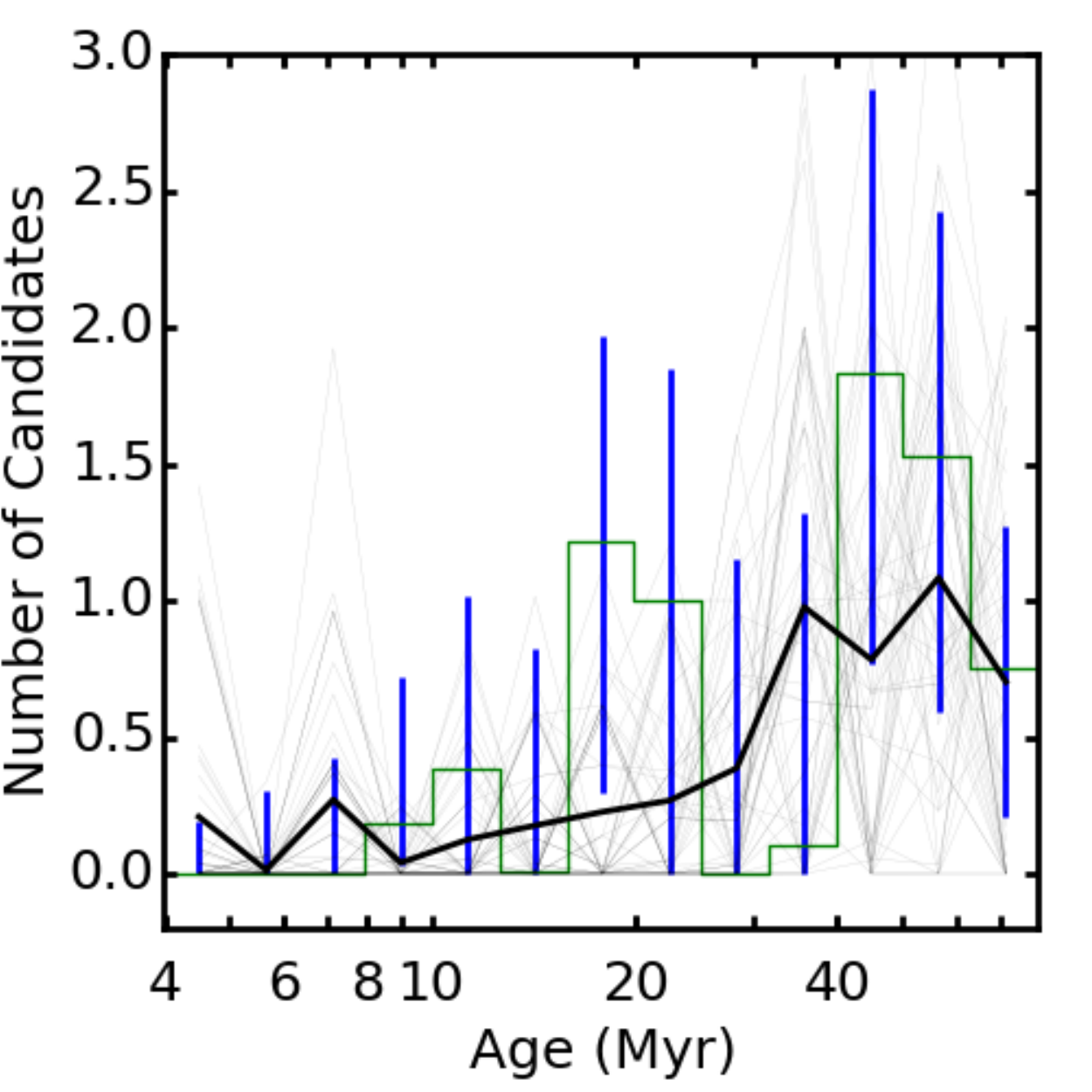}
\caption{Sum of the probability distributions for a few subsamples of the HMXB
  candidates. In each case, the green histogram with blue error bars shows the distribution (and uncertainty) of the candidates, the thin gray lines show the results of measuring the same distribution on equal-sized sets of SFHs taken from positions that do not include an HMXB candidate.  The heavy black line shows the average of all of the thin gray lines.  In short, the black lines show the effects of contamination (young stars that are unrelated to the presence of any HMXBs). {\it Left:} Distribution of all 58 point source candidates with SFH measurements, but only 40 are included because 17 did not reside in a region with recent star formation according to the PHAT data, and one did not have a PHAT SFH. {\it Right:}  A sample of the 7 bluest stellar candidates with SFH measurements, which are our best HMXB candidates, all of which are in regions with recent star formation.}
\label{ages}
\end{figure*}


\begin{figure*}
\begin{center}
\includegraphics[height=4.0in]{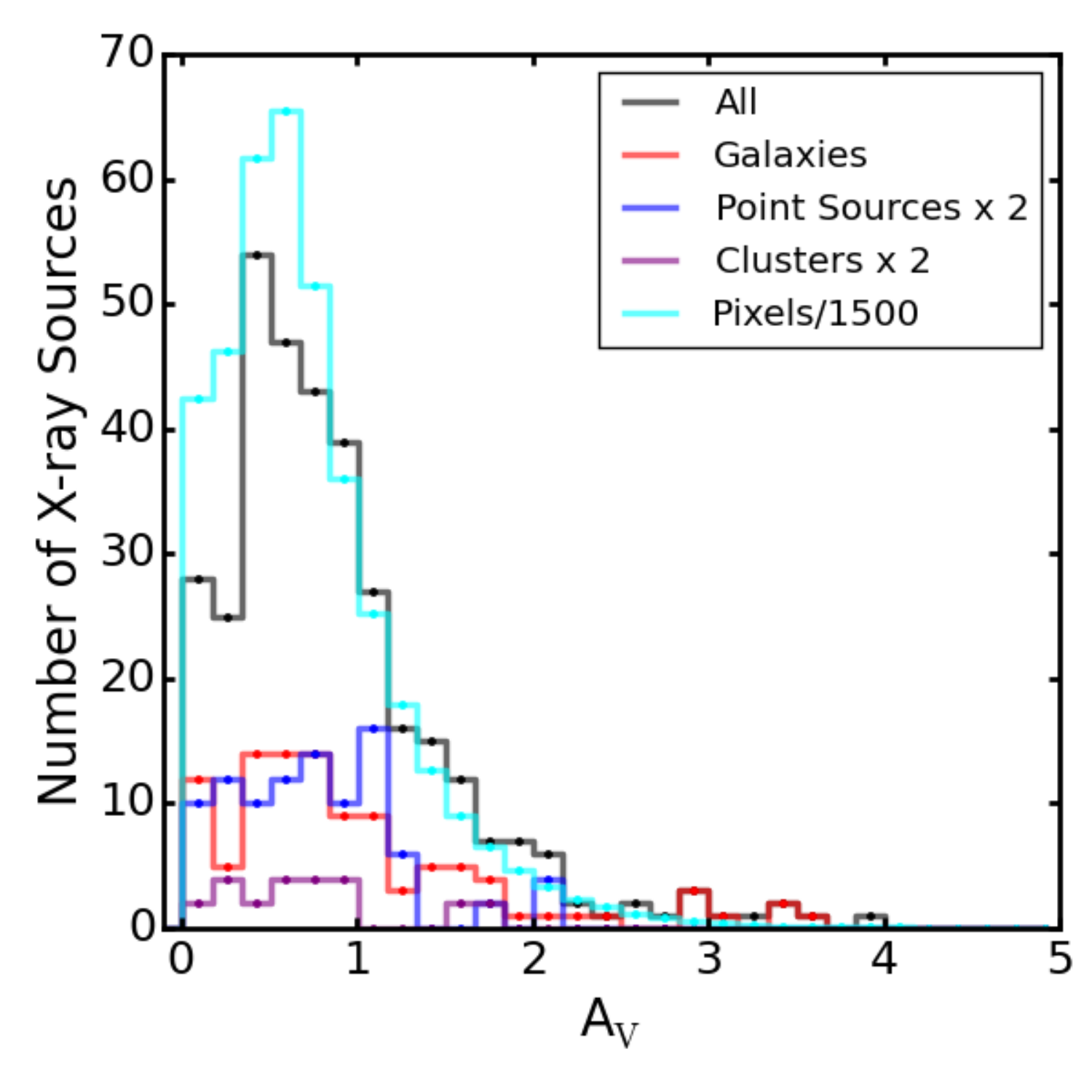}
\end{center}
\caption{Histograms of the mean $A_{\rm V}$ for the entire PHAT survey
  and at the locations of X-ray sources in our survey within the PHAT
  footprint.  Colors in the legend denote quantity being plotted,
  including the distributions of sources with various types of optical
  counterpart candidates. The numbers of point sources and clusters in
  each bin were doubled and the number of pixels in each bin was
  scaled down by a factor of 1500 to make the shapes of their
  distributions visible on the plot.\label{avhist}}
\end{figure*}

\end{document}